\documentclass[twocolumn,english,prb,showpacs,amsmath,amssymb,eqsecnum,preprintnumbers]{revtex4}
\usepackage[T1]{fontenc}
\usepackage[latin9]{inputenc}
\usepackage{color}
\usepackage{bm}
\usepackage{amstext}
\usepackage{esint}
\usepackage{soul}

\makeatletter
\@ifundefined{textcolor}{}
{%
 \definecolor{BLACK}{gray}{0}
 \definecolor{WHITE}{gray}{1}
 \definecolor{RED}{rgb}{1,0,0}
 \definecolor{GREEN}{rgb}{0,1,0}
 \definecolor{BLUE}{rgb}{0,0,1}
 \definecolor{CYAN}{cmyk}{1,0,0,0}
 \definecolor{MAGENTA}{cmyk}{0,1,0,0}
 \definecolor{YELLOW}{cmyk}{0,0,1,0}
}

\def\kF{k_{\text{F}}}
\def\vF{v_{\text{F}}}
\def\NF{N_{\text{F}}}
\def\chiN{\chi_{\text{N}}}
\def\chis{\chi_{\text{s}}}
\def\chin{\chi_{\text{n}}}

\def\epsilonF{\epsilon_{\text{F}}}

\def\qslash{{q\hskip -5pt /}}
\def\Pslash{{P\hskip -6pt /}}
\def\be{\begin{equation}}
\def\ee{\end{equation}}
\def\bea{\begin{eqnarray}}
\def\eea{\end{eqnarray}}
\def\bse{\begin{subequations}}
\def\ese{\end{subequations}}

\def\qslash{q\!\!\!/}
\usepackage{dcolumn}
\usepackage{bm}
\usepackage{amsfonts}
\draft

\makeatother

\usepackage{babel}
\begin{document}

\title{Nonanalyticities in a Strongly Correlated Fermi Liquid: Corrections
to Scaling at the Fermi-Liquid Fixed Point}

\author{D. Belitz$^{1,2}$, T. R. Kirkpatrick$^{3}$}

\affiliation{$^{1}$ Department of Physics and Institute of Theoretical Science,
University of Oregon, Eugene, OR 97403, USA\\
$^{2}$ Materials Science Institute, University of Oregon, Eugene,
OR 97403, USA\\
 $^{3}$ Institute for Physical Science and Technology,and Department
of Physics, University of Maryland, College Park, MD 20742, USA\\
 }

\date{\today}
\begin{abstract}
We use scaling and renormalization-group techniques to analyze the leading
nonanalyticities in a Fermi liquid. We show that a physically motivated scaling
hypothesis reproduce the results known from perturbation theory for the
density of states, the density-of-states fluctuations, the specific heat, the spin 
susceptibility, and the nematic magnetic susceptibility. We also discuss the 
absence of nonanalytic terms in the density susceptibility. We then use
a recent effective field theory for clean electron systems to derive the scaling
hypothesis by means of renormalization-group techniques. This shows that
the exponents (although not the prefactors) of the nonanalyticities that were
previously derived by means of perturbative techniques are indeed exact, and
can be understood as the leading corrections to scaling at the stable
Fermi-liquid fixed point.
\end{abstract}
\maketitle

\section{Introduction, and Results}
\label{sec:I}

\subsection{Introduction}
\label{subsec:I.A}

In any system with soft or massless excitations, the observable behavior
at long wavelengths and low frequencies is dominated by these soft
modes, to the extent that the observables couple to them. In the quantum
regime, the soft modes in addition govern the low-temperature behavior.
In some systems, soft modes exist only at special points in the phase
diagram, for instance at the critical point of a second-order phase
transition. An example are Ising magnets. In others, soft modes exist
in entire phases, either because a spontaneously broken continuous
symmetry leads to Goldstone modes, or because of conservation laws. We will
refer to such soft modes as ``generic''.
An example are the ferromagnons in the ordered phase of a Heisenberg ferrogmagnet. Technically,
scaling ideas and the renormalization group (RG) are well suited to deal
with soft modes and their consequences. In condensed matter physics,
RG techniques are best known for the theory of singularities at critical 
points.\cite{Wilson_Kogut_1974} However,
it arguably is even more interesting and important to understand singularities
that appear in entire phases. The fact that the RG is equally useful for this
purpose is less well known. In the RG
framework, critical singularities are described in terms of critical fixed
points of the RG transformations, whereas entire phases are described
in terms of stable fixed points.\cite{Ma_1976, Cardy_1996} At the former, the
control parameter (in the case of a thermal phase transition, the
temperature deviation from the critical temperature) is the only relevant
operator; at the latter, there are no relevant operators (ignoring external fields
in both cases).

There are many examples of long-wavelength and/or low-frequency
singularities that exist in entire phases due to the coupling of observables to generic
soft modes,\cite{Belitz_Kirkpatrick_Vojta_2005} and we list only a few 
for illustration purposes: (1) The longitudinal 
magnetic susecptibility $\chi_{\text{L}}$ in the ordered phase of an isotropic 
Heisenberg ferromagnet in $d<4$
dimensions diverges as $\chi_{\text L}(h\to 0) \propto h^{-(4-d)}$ as a 
result of the longitudinal magnetization coupling to the magnons.\cite{Brezin_Wallace_1973}
(2) The density of states (DOS) $N$ in a disordered electron system as 
a function of the energy distance $\omega$ from the Fermi energy $\epsilonF$ 
shows a cusp singularity, 
$N(\epsilonF + \omega) \propto \text{const.} + \vert\omega\vert^{(d-2)/2}$, 
as a result of a coupling
to the diffusive modes known as ``diffusons''.\cite{Altshuler_Aronov_1979,
Altshuler_Aronov_1984} In $d\leq 2$ the singularity is so strong that the
disordered Fermi liquid is destroyed. (3) The kinematic viscosity $\nu$ in a classical
fluid is a nonanalytic function of the frequency $\omega$: $\nu(\omega\to 0) \propto
{\rm const.} + \omega^{(d-2)/2}$. This ``long-time tail'' (the corresponding
time correlation function decays as $1/t^{d/2}$ for long times $t$, rather
than exponentially)\cite{Alder_Wainwright_1970, Ernst_Hauge_van_Leeuwen_1970,
Dorfman_Cohen_1970, Pomeau_Resibois_1975} is a consequence of the coupling to various soft modes
in fluids that exist as a result of conservation laws. Again, in $d\leq 2$ the
strong singularity leads to a breakdown of local hydrodynamics. These are just three
examples each of three rather large classes of phenomena in classical
magnets, disordered electrons, and classical fluids, respectively, all of
which can be studied by RG techniques, see Appendixes \ref{app:A},
\ref{app:B}, and \ref{app:C}.

These analogies between classical and quantum systems notwithstanding, there
also are important qualitative differences between them. In classical systems,
the statics and the dynamics are decoupled in equilibrium, meaning that
equilibrium static correlation functions can be long-ranged only due Goldstone
modes. Only in a nonequilibrium situation do the statics and dynamics
couple, and long-range static correlations can result from conservation laws as
well.\cite{Dorfman_Kirkpatrick_Sengers_1994} In quantum systems, on the other hand, the statics and dynamics are
intrinsically coupled even in equilibrium, and long-range static correlations
can result from either Goldstone modes or conservation laws.

In this paper we use RG and scaling techniques
to discuss the interesting singularities that occur in
a clean Fermi liquid in dimensions $d>1$. It is well known that the leading
behavior of interacting electrons at low temperature, frequency, and wave
number is described by Landau's Fermi-liquid theory,\cite{Baym_Pethick_1991}
which corresponds to a stable fixed point in a RG framework.\cite{Shankar_1994}
One major point of the present paper is to show that various nonanalytic 
corrections to Fermi-liquid theory that have been previously derived using perturbation
theory can be understood as the leading corrections to scaling at this stable
fixed point. As we will see, there are strong technical and physical analogies between 
these singularities and the ones in disordered Fermi liquids, and to a lesser extent in
classical magnets and classical fluids, mentioned above. Another important point is that
RG techniques will allow us to show that the perturbative results 
are actually exact as far as the exponents that characterize the
singularities are concerned. In $d=1$ the singularities we will discuss
are so strong that they contribute to an instability of the Fermi liquid in favor
of a Luttinger-liquid state.\cite{Giamarchi_2004} In strongly correlated systems, they
may lead to a quantum phase transition from a Fermi liquid to a non-Fermi liquid state
even in $d>1$.\cite{Kirkpatrick_Belitz_2012}

This paper is organized as follows. In Sec. \ref{sec:II} we formulate a scaling hypothesis
for the free energy. From this we derive homogeneity laws for the observables of interest,
which all can be expressed as derivatives of the free energy. The resulting leading 
nonanalytic corrections to Fermi-liquid theory are all consistent with previous results
from explicit perturbative calculations. In Sec. \ref{sec:III} we recall the schematic form
of a recent effective field theory for clean fermions.\cite{Belitz_Kirkpatrick_2012} We
identify the fixed-point action that describes the Fermi liquid, and identify the leading
irrelevant operators with respect to this fixed point. This allows for a derivation of the
scaling hypothesis, and hence shows that the exponents derived from the scaling
considerations are exact. In Sec.\ \ref{sec:IV} we discuss various aspects of our
approach, and our results. Analogies with classical magnets, disordered electron
systems, and classical fluids, respectively, that are of physical or pedagogical interest
are the subject of Appendices \ref{app:A}, \ref{app:B}, and \ref{app:C}. Appendix \ref{app:D}
explains a structural feature of the density-of-states susceptibility.

\section{Phenomenological scaling considerations for the free energy and its
derivatives}

\label{sec:II}

Before we go into the technical details of
applying RG ideas to analyze a microscopic theory, let us employ simple
phenomenological scaling arguments \cite{scaling_footnote} to find
out what one should expect for the behavior of various thermodynamic
observables, as well as the density
of states and its fluctuations, in a clean Fermi liquid. Let us assign a scale
dimension $[L]=-1$ to lengths, or $[k]=1$ to wavenumbers, and a
scale dimension $[E]=[\omega]=z$ to energy and frequency; that is,
the frequency scales with the wave number as $\omega\sim k^z$.\cite{notation_footnote}
$[L]$ is thus fixed by a convention, whereas the value of $z$ depends
on the nature of the soft modes in the system, and in general there
may be more than one scale dimension $z$, see below. Now
consider an observable $A$ that depends on the wave number $k$, 
the frequency $\omega$, the temperature $T$, and the magnetic field $H$,
whose scale dimension $[H]$ depends on the physical situation.
In general $A$ will consist of a regular or nonscaling part, and a scaling
part $\delta A$ with a scale dimension $[A]$ that is given by the naive
or engineering dimension of $A$.\cite{scale_dimension_footnote} 
The scaling hypothesis states that $\delta A$ obeys a homogeneity law
\be \delta A(k,\omega,T,H) = b^{-[A]}\,\delta A(kb,\omega b^z,Tb^z, Hb^{[H]})\ ,
\label{eq:2.1} 
\ee 
where $b>0$ is the (arbitrary) length rescaling
factor. The nonscaling part does not satify such a homogeneity law. 

If there is more than one class of soft modes in a system, then there will
be more than one dynamical exponent $z$. This is more common at critical
fixed points than at stable ones, and for much of our discussion there will
be only one $z$. However, for electrons interacting via a long-ranged
Coulomb interaction the plasmon is soft for all dimensions $d<3$, and it
scales differently than the other soft modes, so there are two separate
dynamical exponents. As long as the scaling functions in question
are regular functions of their arguments in the limit of small arguments,
the largest of these various $z$ will determine the leading nonanalytic 
behavior of the observable. However, there are important
exceptions to this rule if the scaling function is {\em not} a regular
function of all of its arguments. The DOS in the presence of
plasmons represents an example, as we will see below. 

We finally note that the choice of variables one assigns a nonzero
scale dimension to is physically motivated, and depends on the fixed
point under consideration. Generally, variables that take the system
away from the fixed point under consideration, or take it from one
fixed point to another one with a different symmetry, carry a positive
scale dimension. An example is the magnetic field. By contrast, a change of the
chemical potential $\mu$ takes the system from a Fermi-liquid fixed point to
an equivalent Fermi-liquid fixed point with the same symmetry. The
chemical potential is therefore assigned a scale dimension of zero, $[\mu]=0$,
although its naive dimension is that of an energy. This is a simple, but
important point: The scaling hypothesis involves much more than just
dimensional analysis.

\subsection{Soft modes, and universality classes}
\label{subsec:II.A}

We are interested in universal behavior
at low temperature in the limit of small wave numbers and small frequencies
that is caused by the presence of soft modes. The first question therefore
needs to be about the nature of the soft modes in a Fermi liquid.

There are two types of modes that are soft at $T=0$. The first class
are single-particle excitations represented by the Green function
$\langle{\bar{\psi}}_{1}\,\psi_{2}\rangle$, where $\bar{\psi}_{1}$
and $\psi_{2}$ are fermionic fields with labels $1\equiv({\bm{x}}_{1},\tau_{1},\sigma_{1})$,
$2\equiv({\bm{x}}_{2},\tau_{2},\sigma_{2})$, etc. that comprise position
${\bm{x}}$, imaginary time $\tau$, and spin projection $\sigma$.
These excitations are soft at $T=0$ because of the existence of a
Fermi surface. They have a linear frequency-momentum relation (with the
momentum measured from the Fermi wave number), and are effectively
one-dimensional since only excitations perpendicular to the $(d-1)$ dimensional
Fermi surface are relevant. We will refer to them as the fermionic excitations.
They determine the leading scaling behavior in a Fermi liquid, and
they play a central role in Shankar's RG derivation of Landau Fermi-liquid
theory.\cite{Shankar_1994} 

Much more important for our purposes is a second class 
of soft modes. These are two-particle
excitations of the type $\langle{\bar{\psi}}_{1}\,{\bar{\psi}}_{2}\,\psi_{3}\,\psi_{4}\rangle$.
Since bilinear products of fermionic fields $\psi$ and their adjoints
$\bar{\psi}$ commute with each other as well as with fermion fields,
these are effectively bosonic excitations, and we will refer to them
as such. They in turn fall into two distinct classes. The first one are
the familiar particle-hole-continuum excitations,\cite{Pines_Nozieres_1989}
and the corresponding excitations in the particle-particle or Cooper channel.
They all have a linear frequency-momentum relations, and they appear in all angular 
momentum channels; in the s-wave or $\ell=0$ channel
suitable linear combinations of them constitute the number density, spin density,
and particle-particle density fluctuations. They are responsible, {\it inter alia},
for the familiar structure of the Lindhard function with its linear frequency-momentum
scaling.\cite{Pines_Nozieres_1989} The second class comprises collective
excitations. In the particle-hole spin-singlet channel, these include the excitations known
as zero-sound modes in a neutral system, and the plasmons in a charged system. In the
particle-hole spin-triplet channel, there are the paramagnon excitations. Both the
zero-sound modes and the paramagnons also
have a linear frequency-momentum relation, but their origin and physical
nature is very different from the first class. They are the result of conservation laws
(particle number and spin conservation, respectively) that guarantee their masslessness.
The particle-hole continuum excitations, on the other hand, can be understood as
the Goldstone modes of a spontaneously
broken continuous symmetry that can be represented as a rotational
symmetry between retarded and advanced degrees of freedom.\cite{Belitz_Kirkpatrick_2012}
Their softness is therefore not accidental either; it is controlled by a Ward identity and
protected by Goldstone's theorem.
As we will see, they provide the leading corrections to scaling in
a Fermi liquid, and the effective field theory of Ref.\ \onlinecite{Belitz_Kirkpatrick_2012}
is formulated in terms of them. It is important to note that, due to the intrinsic coupling
between the dynamics and the statics in a quantum system, these bosonic excitations
lead, via mode-mode coupling effects, to long-ranged static correlations even in
equilibrium systems. The same is not true for the fermionic excitations. These long-ranged
correlations in turn fundamentally modify the nature of various quantum phase
transitions.\cite{Belitz_Kirkpatrick_Vojta_2005, Kirkpatrick_Belitz_2011a}

The fact that all of the soft modes mentioned above have linear frequency-momentum
relations is accidental. In systems with a long-ranged Coulomb interaction there is
the plasmon excitation, whose frequency scales as the square root of the wave number
in two-dimensional systems, and as a constant in three-dimensional ones. Another
example of bosonic excitations whose frequency-momentum relation is different from
that of the fermionic ones are the diffusive modes in a disordered electron system.\cite{Altshuler_Aronov_1984}
A technical consequence in all of these cases is the existence of more than on
dynamical scale dimension $z$.

The Goldstone modes in the particle-hole spin-triplet channel and in the particle-particle
channel are sensitive to an external magnetic field. The orbital effects of the field give a
mass to the particle-particle channel, whereas the Zeeman effect gives a mass to two out
of three modes in the particle-hole spin-triplet channel. We therefore have to distinguish
between two universality classes: (1) A generic class with no symmetry-breaking fields,
where all channels are soft, and (2) a magnetic-field universality class, where the Cooper
channel is missing from the soft-mode spectrum, and only one mode in the particle-hole 
spin-triplet channel (the longitudinal one)  is soft.\cite{disordered_universality_classes_footnote}
For the DOS and its susceptibility we further have to distinguish between the cases of a
short-ranged interaction and a long-ranged Coulomb interaction, as we will see in 
Sec.\ \ref{subsec:II.E} below.

\subsection{Scaling of the free energy} 
\label{subsec:II.B}

Let us now consider the free energy density, from which all thermodynamic
quantities can be derived (see below for the relation between the density
of states and the free energy). In principle, theories formulated in terms of
fermionic degrees of freedom only, such as Ref.\ \onlinecite{Shankar_1994}, and
theories formulated in terms of bosonic variables, such as Ref.\ \onlinecite{Belitz_Kirkpatrick_2012},
both contain the effects of both types of soft modes. However,
in practice it is very difficult to extract the effects of one class
of soft modes from a theory that has been formulated in terms of the
other, and the effects of the soft modes in the limit of long wavelengths and low frequencies
are additive. It therefore is natural to postulate the existence
of two additive scaling contributions to the free energy density $f$,
which we denote by $f^{(f)}$ and $f^{(b)}$, respectively, with the
superscripts referring to the fermionic and bosonic soft modes, 
respectively.\cite{scaling_parts_footnote}
Both depend on the temperature $T$ and the magnetic field $H$. In
addition, they depend on a source field $h$ that is conjugate to the density
of states. $h$ is a generalized, frequency
dependent (for simplicity we do not show the frequency dependence explicitly), 
chemical potential, and the DOS can be interpreted as the order parameter of the broken
symmetry.\cite{Kirkpatrick_Belitz_2012, Belitz_Kirkpatrick_2012} For the scaling part of the
free energy density we thus write 
\be 
f(T,H,h) = f^{(f)}(T,H,h) + f^{(b)}(T,H,h)\ , 
\label{eq:2.2} 
\ee 
and we postulate that $f^{(f)}$ and $f^{(b)}$ obey separate homogeneity laws.

We now need to determine the scale dimensions of $f^{(f)}$ and $f^{(b)}$,
and their arguments. To this end we observe that, dimensionally, $f^{(f)}$
and $f^{(b)}$ are both an energy divided by a volume. However, for
the fermionic excitations only the direction perpendicular to the
Fermi surface is relevant,\cite{Shankar_1994} and the scale dimension
of the fermionic part should therefore be 
\bse
\label{eqs:2.3}
\be
[f^{(f)}]=1+z\ .
\label{eq:2.3a}
\ee
For the bosonic part, we expect 
\be
[f^{(b)}] = d+z\ , 
\label{eq:2.3b}
\ee
with $d$ the spatial dimensionality
of the system. The temperature has a scale dimension $[T]=z$, see
above. For both the fermionic and bosonic modes the frequency scales
as the wave number, and we therefore put 
\be
[T] = z = 1\ . 
\label{eq:2.3c}
\ee
We note, however, that in the case of a Coulomb interaction there is a second time scale
set by the plasmon excitation, and hence a second dynamical exponent $z$, see the remarks
in Sec.\ \ref{subsec:II.A} and
after Eq.\ (\ref{eq:2.1}). We will explicitly deal with this in the context of Eqs.\ (\ref{eq:2.11})
and (\ref{eq:2.19}) below. We will not consider orbital
effects of the magnetic field, i.e., we consider $H$ only as it enters
a Zeeman term, and hence 
\be
[H] = 1\ . 
\label{eq:2.3d}
\ee
Finally, $h$ is a generalized
chemical potential and therefore dimensionally an energy, so we also have 
\be
[h] = 1\ . 
\label{eq:2.3e}
\ee
However, the physical chemical potential carries
a scale dimension of zero, see the remark at the end of the introduction
to the current section,
\be
[\mu] = 0\ .
\label{eq:2.3f}
\ee
\ese
With these definitions, the DOS is given by $N=(\partial f/\partial h)/T$,
which correctly makes $N$ an inverse energy times an inverse volume.
Combining all of these considerations, we now have 
\bse 
\label{eqs:2.4}
\bea 
f^{(f)}(T,H,h,\mu) &=& b^{-(1+1)}\, f^{(f)}(Tb,Hb,hb,\mu) \qquad
\label{eq:2.4a}\\
f^{(b)}(T,H,h,\mu) &=& b^{-(d+1)}\, f^{(b)}(Tb,Hb,hb,\mu) \ .
\label{eq:2.4b}
\eea 
\ese

We note that there is an important physical difference between the
physical fields $H$ and $T$, and the derivatives of $f$ with respect to them
(i.e., thermodynamic quantities) on one hand, and the field $h$, which just
serves as a source term that cannot by physically realized, on the other. A
related point is that the derivatives of $f$ with respect to $h$, viz., the
DOS and its susceptibility, are not gauge invariant quantities
and hence show scaling behavior that is sensitive to a long-ranged
Coulomb interaction, whereas the thermodynamic quantities show the
same behavior for both short-ranged and Coulomb interactions. See
Sec.\ \ref{subsec:IV.B} for a discussion of this point.

Before we analyze these homogeneity laws, let us illustrate an alternative
way to determine the effective scale dimension of $f^{(f)}$. Suppose
we do not use the above argument about effectively $1$-$d$ fermionic 
excitations, and assume instead that $f^{(f)}$ scales as an energy divided
by a volume. However, we do acknowledge that the free energy depends
on the microscopic wave number $\kF$ and the microscopic energy $\epsilonF$,
and use as input that $f^{(f)}$ is proportional to $\kF^{d}/\epsilonF$.\cite{Landau_Lifshitz_V_1980}
We thus write 
\bse 
\label{eqs:2.5} 
\be f^{(f)}(T,h,H,\mu) = \frac{\kF^d}{\epsilonF}\,{\cal F}(T,h,H,\mu)\ . 
\label{eq:2.5a} 
\ee 
Then ${\cal F}$ is dimensionally
an energy squared, and the scaling hypothesis is 
\be {\cal F}(T,H,h,\mu) = b^{-2}\,{\cal F}(Tb,Hb,hb,\mu)\ , 
\label{eq:2.5b} 
\ee 
\ese
which is equivalent to Eq.\ (\ref{eq:2.4a}). Knowledge about the
dependence of $f^{(f)}$ on the microscopic parameters is thus equivalent
to knowledge about the effective one-dimensionality of the fermionic
excitations. In Section \ref{subsec:IV.A} we will revisit these arguments
and show how one can write a single free energy contribution that
yields both the fermion and boson contributions by using the
dependence of $f$ on irrelevant operators.

\subsection{Oberservables as derivatives of the free energy}
\label{subsec:II.C}

The observables considered in this paper can be expressed in terms of
derivatives of the free energy as follows. The DOS is given by
the first derivative of $f$ with respect to $h$,
\bse
\label{eq:2.6}
\be
N=\frac{1}{T}\left(\frac{\partial f}{\partial h}\right)_{h=0}\ ,
\label{eq:2.6a}
\ee
and the density-of-states susceptibility by the second one,
\be
\chi_{N}=\frac{1}{T}\left(\frac{\partial^{2}f}{\partial h^{2}}\right)_{h=0}\ .
\label{eq:2.6b}
\ee
\ese
The entropy density $s$ is the first derivative of $f$ with respect to $T$,
\bse
\label{eqs2.7}
\be
s = \partial f/\partial T\ ,
\label{eq:2.7a}
\ee
and the specific-heat coefficient $\gamma_V = C_V/T$ is the second one,
\be
\gamma_V = \partial^2 f/\partial T^2\ .
\label{eq:2.7b}
\ee
\ese
The magnetization $m$, defined as the magnetic moment per unit volume, 
is the first derivative of $f$ with respect to $H$,
\bse
\label{eqs:2.8}
\be
m = \partial f/\partial H\ ,
\label{eq:2.8a}
\ee
and the spin susceptibility is the second one,
\be
\chi_{\text{s}} = \partial^2 f/\partial H^2\ .
\label{eq:2.8b}
\ee
Note that differentiating with respect to $H$ yields the spin susceptibility, rather
than the full magnetic susceptibility, since we have restricted the $H$-dependence
of the Hamiltonian to a Zeeman term.
Also of interest is the (p-wave) nematic spin susceptibility that is given as the
second derivative with respect to a field ${\cal H}$ that couples to the spin
current rather than to the spin density,
\be
\chi_{\text{s,p-wave}} = \partial^2 f/\partial{\cal H}^2\ .
\label{eq:2.8c}
\ee
\ese
Finally, to make a point about naive dimensions and scaling functions,
we also consider the density susceptibility,
\be
\chin = \partial n/\partial\mu = \partial^2 f/\partial\mu^2\ ,
\label{eq:2.9}
\ee
with $n$ the particle number density. For reasons discussed below, 
$\chin$ scales differently than the other susceptibilities.

All of the above quantities have a naive dimension of inverse volume times
inverse energy, except for $\chiN$, which is an inverse volume times an
inverse energy squared. We will eventually conclude that in the case of
short-ranged electron interaction, where there is only one dynamical
exponent $z$, all of them except for $\chin$ have scaling properties
that are simply given by their naive dimension. $\chin$, however, does
not scale according to its naive dimension. The physical reason is that
changing the chemical potential does not change the symmetry of the
system, nor does it move the system away from the stable, zero-temperature
Fermi-liquid fixed point. That is, $\mu$ is not a RG-relevant operator
with respect to the Fermi-liquid fixed point. This conclusion is confirmed
by explicit perturbative calculations.\cite{Belitz_Kirkpatrick_Vojta_1997}

\subsection{The leading Fermi-liquid behavior}
\label{subsec:II.E}

The formulation of Fermi-liquid theory in terms of a stable RG fixed point
has been studied in great detail.\cite{Shankar_1994, Nayak_Wilczek_1996} 
Here we give some simple arguments to show that our treatment is consistent 
with the existence of a stable Fermi-liquid fixed point, but we will not deal
with the full complexity of the Fermi-liquid state. Our main objective will be the study of the
leading corrections to scaling at this fixed point, which result from the
bosonic soft modes. In this context, the Fermi-liquid results appear as a
regular background for the nonanalyticities due to the bosonic soft modes
and the resulting long-range correlations. 

Let us consider the contributions of the fermionic excitations to the
various observables. From Eqs.\ (\ref{eq:2.6a}) and (\ref{eq:2.4a}) or, alternatively,
(\ref{eqs:2.5}), we see that the DOS scales as a constant, $N \sim \text{const}$.
Putting $b=1/T$ in Eq.\ (\ref{eq:2.4a}) we see that the entropy density scales
as $s \sim T$, and the specific-heat coefficient scales as $\gamma_V \sim \text{const}$.
For the magnetization, putting $b=1/H$ in Eq. (\ref{eq:2.4a}) yields $m \sim H$,
and hence the spin susceptibility also scales as a constant, $\chis \sim \text{const}$.
The same is true for the nematic susceptibility $\chi_{\text{s,p-wave}}$ (and all higher
nematic susceptibilities as well). 

Using Eq.\ (\ref{eq:2.9}) and either 
Eq.\ (\ref{eq:2.4a}) or (\ref{eq:2.5b}), we see that $\chin \sim \text{const}$ as well.
The scaling assumption thus reproduces the well-known properties of a Fermi liquid: 
The DOS and the susceptibilities $\gamma_V$, $\chis$, and $\chin$ all are propotional to the 
bare DOS at the Fermi surface, $\NF$. Notice, however, that the origin
of $\chin$ scaling as a constant is very different from the analogous statements for
the other susceptibilities, as it is a consequence of $[\mu]=0$. The fact that all of
these susceptibilities are trivially proportional to $\NF$ by dimensional analysis
thus masks an important difference. We will come back to this point below. The fields
$H$ and $h$ do not couple to the fermionic excitations in any interesting
way, and the latter therefore do not produce any nonanalytic corrections
to the leading behavior.

Finally, for the DOS susceptibility $\chiN$, Eqs.\ (\ref{eq:2.4a}) and (\ref{eq:2.6b}) suggest
$\chiN \sim 1/T$. This is misleading, however. Repeated differentiations with respect
to the conjugate field $h$ at the same frequency do not lead to any frequency mixing,
and as a consequence the Fermi-liquid result for $\chiN$ is not as singular as naive
scaling suggests; the second differentiation with respect to $h$ in Eq.\ (\ref{eq:2.6b})
just produces a factor of $1\epsilonF$. In fact, the wave-number ($k$) and frequency ($\omega$) dependent generalization
of $\chiN$ at $T=0$ for free electrons is $\chiN^{(0)} \propto \NF/\epsilonF$.
Fermi-liquid corrections do not change the scaling of this result, and therefore we have
$\chiN \sim \text{const}$. As we will see in the next subsection, for this quantity the 
leading contribution from the bosonic fluctuations in $d\leq 2$ is actually {\em stronger} than the Fermi-liquid contribution.

\subsection{The leading corrections to Fermi-liquid scaling}
\label{subsec:II.E}

Now consider the contributions of the bosonic excitations to the scaling
part of the free energy, which yield the leading nonanalytic dependence
of variables observables on the frequency, temperature, etc., and are
not included in Landau Fermi-liquid theory. We will discuss the DOS,
its susceptibility, the specific-heat coefficient, and the magnetic susceptibility, 
and distinguish between the cases of a short-range interaction
and a long-range Coulomb interaction as appropriate. We note that the
leading corrections to scaling can be large effects that are very important
for an understanding of strongly correlated metals, and even have the potential
for destroying the underlying Fermi-liquid state.\cite{Kirkpatrick_Belitz_2012}

\subsubsection{The density of states}
\label{subsubsec:II.E.1}

The leading correction to the constant Fermi-liquid DOS is given by
$\delta N = (\partial f^{(b)}/\partial h)/T\vert_{T=h=0}$. Let us first
consider the case of a short-ranged interaction, in which the complications
due to the presence of multiple time scales mentioned above are not
relevant.

\paragraph{Short-ranged case}
\label{par:II.E.1.a}

From Eq.\ (\ref{eq:2.4b}) we have 
\bse 
\label{eqs:2.10} 
\be 
\delta N(\omega, T) = b^{1-d}\,\delta N(\omega b, T b)\ , 
\label{eq:2.10a}
\ee 
This implies 
\be 
\delta N(\omega, T=0) \propto \vert\omega\vert^{d-1}
\label{eq:2.10b} 
\ee 
and
\be
\delta N(\omega=0,T) \propto T^{d-1}\ ,
\label{eq:2.10c}
\ee
and more generally
\be
\delta N(\omega, T) = \vert\omega\vert^{d-1}\,F_N(T/\omega)\ ,
\label{eq:2.10d}
\ee
\ese
where $F_N(x) = \delta N(1,x)$ is a scaling function.
That is, the DOS at $T=0$ is a nonanalytic function of the energy distance $\omega$ from
the Fermi surface, and the DOS at the Fermi surface is a nonanalytic function of
the temperature. For $1<d<3$ the leading nonanalyticity is
stronger than the leading analytic correction, which is $\omega^2$ or $T^2$, respectively.
Note that scaling by itself does not guarantee that the prefactor of the
nonanalyticity is nonzero in any given system in any given dimension.
Also, the presence of dangerous irrelevant variables may invalidate
the simple scaling assumption represented by Eq.\ (\ref{eq:2.10a}).\cite{Ma_1976}
However, barring such exceptional circumstances, the exponent in Eq.\ (\ref{eq:2.10b})
is expected to be exact. This is in sharp contrast to perturbative
considerations, which can never guarantee that a stronger nonanalyticity
will not appear at some higher order of perturbation theory. The remaining
question is the validity of the scaling assumption. Establishing this
will be the purpose of Sec.\ \ref{sec:III}, where we will derive
the scaling behavior from renormalization-group arguments. This will
establish that the exponent in Eq.\ (\ref{eq:2.10b}) is indeed exact.
The determination of the prefactor requires an explicit calculation. In
Ref.\ \onlinecite{us_tbp} we will present a one-loop calculation that is
not perturbative in the interaction strength and shows that
the prefactor is generically nonzero. However, the
case $d=2$ is an exception; to one-loop order the prefactor of the
$\vert\omega\vert$ vanishes, in agreement with previous results from
many-body perturbation theory.\cite{Mishchenko_Andreev_2002, Chubukov_et_al_2005}
The leading nonanalytic contribution at one-loop order for $d=2$ is an $\vert\omega\vert/\ln^{3}\vert\omega\vert$
that originates from the particle-particle channel.\cite{higher_loop_footnote} The explicit loop
expansion also shows that in special dimensions there are logarithmic
corrections to power-law scaling, which the scaling theory is not
sensitive to. For instance, in $d=3$ we find $\delta N\propto\omega^{2}\log\vert\omega\vert$.

\paragraph{Long-range case}
\label{par:II.E.1.b}

As we will see in Secs. \ref{sec:III} and \ref{sec:IV}, the simple
scaling arguments given above do indeed yield the correct results
in the case of a short-ranged interaction between the electrons. 
However, in the case of a long-ranged Coulomb interaction in $d\leq 2$ the leading
singularity of the DOS is even stronger. The
reason is the presence of the plasmon time scale, whose frequency
scales as $\omega\sim k^{(3-d)/2}$. In addition to the frequency
scale with $z=1$, which reflects the particle-hole excitations, we
thus have a second frequency scale with $z = z_p = (3-d)/2$. As a result,
Eq.\ (\ref{eq:2.10a}) gets generalized to 
\bea 
\delta N(\omega) &=& b^{1-d}\,\delta N(\omega b,\omega b^{(3-d)/2}) 
\nonumber\\
&=& \vert\omega\vert^{d-1}\,f_N(\omega^{(d-1)/2})\ , 
\label{eq:2.11}
\eea 
where $f_N(x) = \delta N(1,x)$ is a scaling function and we have dropped
the temperature dependence for simplicity. Naively, one would expect
$f_{N}(x\to 0)  =\text{const.}$, which would lead to Eq.\ (\ref{eq:2.10b}).
However, it turns out that the subleading frequency scale characterized
by $z_p$ is a dangerous irrelevant variable for the DOS if $d<2$,
namely, $f_{N}(x\to0)\propto x^{-(2-d)/(3-d)}+O(1)$. As a result,
\be 
\delta N(\omega) \propto \begin{cases} \vert\omega\vert^{(d-1)/(3-d)} & \text{for $1<d<2$}\\
                                                                \vert\omega\vert^{d-1}             & \text{for $2\leq d<3$}\ . 
                                         \end{cases} 
\label{eq:2.12}
\ee
Note that the existence of a dangerous irrelevant variable cannot
be deduced from scaling arguments alone; establishing the behavior
described above requires an explicit calculation.\cite{Mishchenko_Andreev_2002,
Chubukov_et_al_2005} In $d=3$ the leading behavior is again $\omega^2\ln\vert\omega\vert$,
as in the short-range case.

\subsubsection{The density-of-states susceptibility}
\label{subsubsec:II.E.2}

We next recall what scaling arguments predict for the temperature
and frequency-dependence of the density-of-states fluctuations.\cite{Kirkpatrick_Belitz_2013}
Again, we first consider the short-range case.

\paragraph{Short-range case}
\label{par:II.E.2.a}

Let us generalize Eq. (\ref{eq:2.10a}) by keeping the auxiliary source field $h$ conjugate
to the DOS. We then obtain 
\be 
\delta N(\omega, T, h) = b^{(1-d)}\,\delta N(\omega b, T b, h b)\ . 
\label{eq:2.13} 
\ee
Differentiating again with respect to $h$, and putting $h=0$, we obtain a homogeneity
laws for the DOS susceptibility, 
\be \chiN(k,\omega,T) = b^{2-d}\,\chiN(kb, \omega b, Tb)\ ,
\label{eq:2.14}
\ee
where we have added a dependence on the wave number $k$. Alternatively, we can
consider the homogeneous susceptibility in a system with a finite linear dimension $L$.
All homogeneity laws then remain valid with $k$ replaced by $1/L$. Since frequency
and temperature scale the same way, we put $\omega=0$ for simplicity. For the
static susceptibility as a function of $k$ and $T$ Eq.\ (\ref{eq:2.14}) implies
\be \chiN(k, T) = k^{d-2}\,f_{\chi}(T/\vF k)\ , 
\label{eq:2.15} 
\ee 
with $f_{\chi}$ a scaling function and $\vF$ the Fermi velocity. At this point we need
to acknowledge that $\chiN$, because of the frequency structure of the underlying
four-fermion correlation function, has two intrinsically different parts, one of which
is constant as $T\to 0$, whereas the other vanishes linearly with $T$. This is explained
in Appendix \ref{app:D}, and it is consistent with general statistical arguments, see
Ref.\ \onlinecite{Kirkpatrick_Belitz_2013} and Sec. \ref{subsec:IV.E} below. 
$f_{\chi}$, and hence $\chiN$, therefore come with two scaling parts, viz.
\bse
\label{eqs:2.16}
\bea
\chiN^{(0)}(k,T) &=& \frac{1}{k^{2-d}}\,f_{\chi}^{(0)}(T/\vF k)\ ,
\label{eq:2.16a}\\
\chiN^{(1)}(k,T) &=& \frac{T}{k^{3-d}}\,f_{\chi}^{(1)}(T/\vF k)\ ,
\label{eq:2.16b}
\eea
\ese
with $f_{\chi}^{(0)}(x\to 0) = \text{const.}$ and $f_{\chi}^{(1)}(x\to 0) = \text{const.}$ 
In the limit $T\ll\vF k$, $\chiN^{(1)}$ is thus small compared to $\chiN^{(0)}$ by a factor
of $T/\vF k$. In the opposite limit, $\vF k\ll T$, both parts yield
\be
\chiN(k=0,T) \propto 1/T^{2-d}\ .
\label{eq:2.17}
\ee
Of particular interest are the physical dimensions $d=2$ and $d=3$, where the zero 
exponents in Eqs.\ (\ref{eqs:2.16}, \ref{eq:2.17}) signify logarithms. An explicit calculation\cite{us_tbp}
yields for the leading behavior in $d=2$
\bse
\label{eqs:2.18}
\be
\chiN(k,T) \propto \begin{cases} \ln(1/k) & \text{for $T\ll\vF k$} \\
                                                             \ln(1/T) & \text{for $\vF k\ll T$}
                                      \end{cases}
\label{eq:2.18a}
\ee
from $\chiN^{(0)}$, and in $d=3$,
\be
\chiN(k,T) \propto \begin{cases} k\left[1 + (T/\vF k)\ln(1/k)\right] & \text{for $T\ll\vF k$} \\
                                                     T\ln(1/T) & \text{for $\vF k\ll T$}
                                      \end{cases}\ ,
\label{eq:2.18b}
\ee
\ese
with the logarithms coming from $\chiN^{(1)}$.

This highly nonanalytic behavior of the DOS susceptibility reflects the
same correlations that lead to the nonanalyticity in the DOS itself,
Eq.\ (\ref{eq:2.10b}). Again, the exponents are expected to be exact.

\paragraph{Long-range case}
\label{par:II.E.2.b}

Similar to the case of the DOS, the DOS susceptibility gets modified
by a Coulomb interaction, but for this observable the plasmon frequency
scale is a dangerous irrelevant variable even for $d>2$. Equation\ (\ref{eq:2.15})
gets generalized to 
\bea 
\chiN(k,\omega=0,T) &=& k^{d-2} f_{\chi}\left(\frac{T}{\vF k},\frac{T}{\vF \kappa^{(d-1)/2}\,
                                                                              k^{(3-d)/2}}\right)
\nonumber\\
 &=& k^{d-2} g_{\chi}(T/k,(k/\kappa)^{(d-1)/2})\ , 
\label{eq:2.19}
\eea 
where $\kappa$ is the screening wave number and $g_{\chi}$
is another scaling function. $g_{\chi}(x,y)$ again has two separate scaling
parts, for the reasons explained above. In the long-range case, the one proportional
to $T$ always gives the leading contribution, which generalizes Eq.\ (\ref{eq:2.16b}):
\be
\chiN(k,T) = \frac{T}{k^{3-d}}\,g_{\chi}^{(1)}(T/k,(k/\kappa)^{(d-1)/2})\ .
\label{eq:2.20}
\ee
If $g_{\chi}^{(1)}(x,y)$
were a regular function of its second argument for $y\to 0$, then
the leading behavior of $\chi_{N}$ would again be given by Eq.\ (\ref{eq:2.16b}),
since the plasmon frequency scale is subleading compared to the ballistic
scale, and thus irrelevant in the renormalization-group sense. However,
this subleading scale is again dangerously irrelevant. In order to see this one
needs to perform an explicit calculation that will be reported elsewhere;\cite{us_tbp}
here we just list the results. The dimensionality dependence is complicated, and we focus 
on the physical dimensions $d=2,3$. In $d=2$ one finds\cite{Kirkpatrick_Belitz_2013}
\begin{widetext}
\be
\chiN(k,T) \propto \kappa^2 T \times\begin{cases} (1/T^3)\,\ln(T^2/\vF^2\kappa k) & \text{for $\vF k \ll T^2/\vF\kappa$}\\
                                                     1/(\vF\kappa)^{3/2} T^{3/2}     & \text{for $T^2/\vF\kappa \ll \vF k \ll T$}\\
                                                     1/(\vF\kappa)^{3/2} (\vF k)^{3/2} & \text{for $T\ll \vF k$}\ .
                              \end{cases}
\label{eq:2.21}
\ee
\end{widetext}
In $d=3$ the result is
\be
\chiN(k,T) \propto \kappa^3 T/(\vF\kappa)^2 \vF k
\label{eq:2.22}
\ee
for all values of $T$ and $k$. Note that $\chi_N$ diverges for $k\to 0$ at
fixed $T$ in both $d=2$ and $d=3$. We will discuss the significance of
this result in Sec.\ \ref{sec:IV}.

\subsubsection{The specific-heat coefficient}
\label{subsubsec:II.E.3}

We next consider the specific heat coefficient. The homogeneity
law for the leading correction to the constant Fermi-liquid contribution,
$\delta\gamma_V = \partial^2 f^{(b)}/\partial T^2$, is 
\be 
\delta\gamma_V(T,H) = b^{-(d-1)}\,\delta\gamma(Tb,Hb)\ . 
\label{eq:2.23} 
\ee 
In a zero field, this results in 
\bse \label{eqs:2.24} 
\be 
\delta\gamma_V(T) \propto T^{d-1}\ . 
\label{eq:2.24a} 
\ee 
In a nonzero magnetic field, this gets generalized to 
\be \delta\gamma_V = T^{d-1}\,g_{\gamma}(H/T)\ ,
\label{eq:2.24b} 
\ee 
\ese
with $g_{\gamma}$ a scaling function.
These results are all consistent with perturbation 
theory.\cite{Chubukov_et_al_2005,Betouras_Efremov_Chubukov_2005}

\subsubsection{The spin susceptibility}
\label{subsubsec:II.E.4}

Now we consider the leading correction to the spin susceptibility,
$\delta\chis=\partial^{2}f^{(b)}/\partial H^{2}$. This is due to
soft modes in the spin-triplet channel, and thus is the same for the
short-range and long-range cases. From Eq.\ (\ref{eq:2.4b}), and
including a wave-number dependence of $\chis$, we find 
\be 
\delta\chis(k,T,H) = b^{-(d-1)}\,\delta\chis(kb,Tb,Hb)\ . 
\label{eq:2.25} 
\ee
For the wave-number dependent spin susceptibility at $T=0$ and $H=0$
this yields 
\be \delta\chis \propto k^{d-1}\ . 
\label{eq:2.26}
\ee 
This is consistent with perturbative
results,\cite{Belitz_Kirkpatrick_Vojta_1997, Chitov_Millis_2001, Chubukov_Maslov_2003}
but the scaling arguments are not sensitive to a logarithmic term in $d=3$, where the behavior
is $k^2\ln k$.\cite{Belitz_Kirkpatrick_Vojta_1997} The nonanalytic
$T$ and $H$-dependences of the homogeneous spin susceptibility are
\bse 
\label{eqs:2.27} 
\be 
\delta\chis \propto T^{d-1} 
\label{eq:2.27a}
\ee
and 
\be \delta\chis \propto H^{d-1} 
\label{eq:2.27b} 
\ee
\ese 
respectively. These scaling results are also consistent with
perturbation theory.\cite{Belitz_Kirkpatrick_Vojta_1997, Betouras_Efremov_Chubukov_2005,
Maslov_Chubukov_Saha_2006}

The scaling behavior of $\chi_{\text{s,p-wave}}$ is the same as that of $\chis$, except that
$H$ gets replaced by ${\cal H}$. This is important in the context of the quantum phase
transition from a paramagnet to a spin-nematic phase, and renders the transition generically
first order.\cite{Kirkpatrick_Belitz_2011a} We also note that
$\chi_{\text{s,p-wave}}$ is a nonanalytic function of ${\cal H}$, but not of $H$, since 
${\cal H}$ does not couple to the spin density. Analogous statements hold for other 
susceptibilities and fields: Scaling arguments give only the functional dependence of
observables on fields; a nonzero prefactor requires, among other things, a nonvanishing coupling.

\subsection{The signs of the leading corrections}
\label{subsec:II.F}

The scaling arguments presented in this section give information about the
various power laws that characterize the leading nonanalyticities, but make
no statement about the prefactors. However, with some additional physical
reasoning one can give strong arguments at least for what the signs of the 
various effects should be, which then can be confirmed by explicit calculations.\cite{us_tbp}

Let us start with the spin susceptibility. Its nonanalyticity is a result of fluctuations
about Stoner theory.  This will weaken the
tendency towards ferromagnetism. As a result, $\chis(k=0,T=0)$ will decrease.
A nonzero $k$ or $T$ weakens the soft-mode effect, therefore the sign of the
leading nonanalyticity in Eqs.\ (\ref{eq:2.26}) and (\ref{eq:2.27a}) will be positive. 
This is indeed the result first obtained in
perturbation theory in Ref.\ \onlinecite{Belitz_Kirkpatrick_Vojta_1997}, which has
profound consequences for the ferromagnetic quantum phase 
transition.\cite{Belitz_Kirkpatrick_Vojta_1999, Belitz_Kirkpatrick_Vojta_2005}
We note that in disordered systems the signs of the corresponding nonanalyticities
are opposite.\cite{Kirkpatrick_Belitz_1996} The reason is that disorder slows down
the electrons (diffusive rather than ballistic motion), which effectively enhances
the interaction and hence the tendency towards magnetism. This effect in turn
is weakened by a nonzero $k$ or $T$.
The prefactor in Eq.\ (\ref{eq:2.27b}) is also positive, since a magnetic field
enhances the tendency toward magnetism. This is also in agreement with prior
perturbative results.\cite{Betouras_Efremov_Chubukov_2005}

The signs of the corrections to the DOS and the specific-heat coefficient can be
understood as follows. The correlations induced by the Goldstone modes lead
to long-range correlations that tend to order the system. The entropy is thus
expected to decrease as a result of them, and so will the specific heat. We thus
expect the prefactor in Eqs.\ (\ref{eqs:2.24}) to be negative, which is indeed borne
out by explicit calculations.\cite{Chubukov_et_al_2005,Betouras_Efremov_Chubukov_2005}
This is consistent with the sign of the DOS correction, Eq.\ (\ref{eq:2.10b}), which is 
also known to be negative.\cite{Khveshchenko_Reizer_1998, Mishchenko_Andreev_2002}

\section{RG-Based Derivation of Scaling}
\label{sec:III}

In this section we show how the above results can be derived without invoking a scaling
assumption, by performing a RG analysis of the effective field theory
of Ref. \ \onlinecite{Belitz_Kirkpatrick_2012}. We stress again that even
though the Fermi-liquid fixed point is not a critical fixed point, it nevertheless displays scale
invariance due to the existence of Goldstone modes. Therefore, very
useful results for the entire Fermi-liquid phase can be obtained from very simple
RG arguments. Furthermore, in a properly formulated theory of a stable
phase, non-Gaussian terms are RG irrelevant and, as a consequence of
this, {\em exact} scaling exponents can be simply obtained. This is in contrast
to the situation at a critical fixed point, where the explicit calculation of exponents
usually involves an expansion in an artificial small parameter, such as the
deviation from a critical dimension.\cite{Wilson_Kogut_1974}

\subsection{The structure of the field theory}
\label{subsec:III.A}

The theory of Ref.\ \onlinecite{Belitz_Kirkpatrick_2012} is formulated in term 
of a soft matrix field $q_{nm}({\bm{k}})$ and a massive one $P_{nm}({\bm k})$,\cite{density_formulation_footnote}  
which encode the soft and massive components of bilinear fermion fields $\bar{\psi}_{n}\psi_{m}$,
i.e., those products with $nm<0$ and $nm>0$, respectively. The softness of the $q$ is guaranteed by
a Ward identity. The effective action ${\cal A}$ takes the form of an expansion in powers of $q$ and $P$,
see Eqs.\ (4.45)-(4.47) in Ref.\ \onlinecite{Belitz_Kirkpatrick_2012}.\cite{lambda_footnote}
In a symbolic notation that shows only quantities that carry a scale
dimension, viz., the fields $q_{nm}({\bm{k}})\equiv q$ and $P_{nm}({\bm k}) \equiv P$, and factors
of volume $V$, wave number $k$, and frequency $\omega$ (which we
do not need to distinguish from factors of temperature for our purposes),
the Gaussian action takes the form (see Eq.\ (4.45) in Ref.\ \onlinecite{Belitz_Kirkpatrick_2012})
\be
{\cal A}^{(2)} = \frac{1}{V}\sum_{k,\omega} [k + \omega + \gamma\,\omega]\, q^2 
                        + \frac{1}{V} \sum_{k,\omega} \left[1 + \gamma\omega\right] P^2
\label{eq:3.1}
\ee
Here and in what follows the sums are over the appropriate sets of wave vectors
and frequencies, and the powers of $k$ and $\omega$ in each term
follow from the properties of the convolutions of Green's functions
that make up the vertices of the theory in the limit of long wavelengths
and small frequencies, see Ref.\ \onlinecite{Belitz_Kirkpatrick_2012}.
As mentioned above, $\omega$ can stand for either frequency or temperature. 
The factors $k$ and $\omega$ in the vertices should be understood as being 
multiplied by functions of $k/\omega$, which are of $O(1)$ for scaling purposes 
and are not shown for simplicity. $\gamma$ represents the interaction amplitude.
Notice that the interacting and noninteracting parts of the Gaussian $q$-vertex
both are linear in $k$ or $\omega$, whereas the interacting part of the $P$-vertex
carries a factor of $\omega$ compared to the noninteracting one.

The non-Gaussian terms $\Delta{\cal A}$ are given in terms of fields $\qslash$ and $\Pslash$ that
are closely related to $q$ and $P$. Their operational definition is that their propagators
are the $q$ and $P$-propagators, respectively, with the noninteracting parts 
subtracted.\cite{lambda_footnote} From Eq.\ (\ref{eq:3.1}) we see that the $q$ and $\qslash$
propagators scale the same way, viz.,
\bse
\label{eqs:3.2}
\be
\langle q\,q\rangle \sim \langle\qslash\,\qslash\rangle \sim \frac{V}{k + \omega}\ ,
\label{eq:3.2a}
\ee
whereas the $P$ and $\Pslash$ propagators scale differently,
\bea
\langle P\,P\rangle &\sim& V\times{\text{const.}}
\label{eq:3.2b}\\
\langle \Pslash\,\Pslash\rangle &\sim& V\omega\ .
\label{eq:3.2c}
\eea
\ese
The mixed propagators $\langle q\,\qslash\rangle$ and $\langle P\,\Pslash\rangle$ are equal
to $\langle q\,q\rangle$ and $\langle P\,P\rangle$, respectively.
For scaling purposes we therefore need to distinguish between $P$ and $\Pslash$, but not
between $q$ and $\qslash$. With this in mind, $\Delta{\cal A}$ takes the form (see 
Eqs.\ (4.47, 4.48) in Ref.\ \onlinecite{Belitz_Kirkpatrick_2012})
\be
\Delta{\cal A} = \Delta{\cal A}^{(3)} + \Delta{\cal A}^{(4)} + \ldots
\label{eq:3.3}
\ee
where, in the same schematic notation as in Eq.\ (\ref{eq:3.1}),
\begin{widetext}
\bse
\label{eqs:3.4}
\bea
\Delta{\cal A}^{(3)} &=& \frac{c_{3,0}}{V^2}\sum_{\{k,\omega\}} \left[\gamma\,\omega + O(\gamma^3\omega^3)\right] q^3
   + \frac{c_{2,1}}{V^2} \sum_{\{k,\omega\}} \left[1 + O(\gamma^2\omega^2)\right] q^2 \Pslash 
   + \frac{c_{1,2}}{V^2} \sum_{\{k,\omega\}} \gamma\,\omega\, q \Pslash^{\,2}
   + \frac{c_{0,3}}{V^2} \sum_{\{k,\omega\}} \Pslash^{\,3} 
\label{eq:3.4a}\\
\Delta{\cal A}^{(4)} &=& \frac{c_{4,0}}{V^3}\sum_{\{k,\omega\}} \left[k + \omega + \gamma^2\omega^2/k + O(\gamma^4\omega^4)\right] q^4 
   + \frac{c_{3,1}}{V^3} \sum_{\{k,\omega\}} \left[\gamma\omega/k + O(\gamma^3\omega^3)\right] q^3 \Pslash
\nonumber\\
   &&+ \frac{c_{2,2}}{V^3} \sum_{\{k,\omega\}} \left[1/k + O(\gamma^2\omega^2)\right] q^2 \Pslash^{\,2}
   + \frac{c_{1,3}}{V^3} \sum_{\{k,\omega\}} \gamma\omega\, q \Pslash^{\,3}
   + \frac{c_{0,4}}{V^3} \sum_{\{k,\omega\}} \Pslash^{\,4} \ ,
\label{eq:3.4b}
\eea
\ese
\end{widetext}
where the $c_{n,m}$ are coupling constants.
Notice that various vertices in $\Delta{\cal A}$, e.g., the leading
term of $O(q^2\Pslash^{\,2})$, are singular functions of $k$ (or $\omega$) for small $k$. These
singularities get stronger with increasing order in the fields; for
instance, the leading term of $O(q^2 \Pslash^{\,2n})$ has a vertex that scales as $1/k^{2n-1}$. As we will
show below, these singular vertices do not pose a problem for our purposes. We
stress again that this form of the action is highly schematic and can be used for
power-counting purposes only; many features that are crucial of explicit calculations
have been suppressed for clarity. See Ref.\ \onlinecite{Belitz_Kirkpatrick_2012} for a
complete expression. We also note that the schematic notation ignores a structural
difference between the particle-hole and particle-particle channels that is only
logarithmic in nature and hence does not appear at the level of power counting.
However, it is of qualitative importance once logarithmically small effects are taking into
account, see the next subsection.

\subsection{Leading scaling behavior, and the fixed-point action}
\label{subsec:III.B}

Equation (\ref{eq:3.1}) accurately represents the schematic form of the Gaussian action in the
particle-hole channel, which was the only one considered in Ref.\ \onlinecite{Belitz_Kirkpatrick_2012}.
In particular, it accurately represents the Gaussian propagator in the particle-hole
channel, which has the schematic structure
\be
\langle q\,q\rangle_{\text{p-h}} = \frac{V}{k + (1+\gamma)\omega}\ .
\label{eq:3.5}
\ee
We will proceed by first analyzing the action in the particle-hole channel from
a RG point of view, and then consider the particle-particle or Cooper channel. We also recall
that the particle-particle channel is sensitive to a small magnetic field, which gives
its soft modes a mass, and therefore can always be suppressed, see Sec.\ \ref{subsec:II.A}. 
This effect is qualitatively the same as in disordered electron systems, where the orbital effects
of a small magnetic field suppress the diffusive modes known as 
Cooperons.\cite{Lee_Ramakrishnan_1985} 

We now look for a fixed point of the action, Eqs.\ (\ref{eq:3.1}, \ref{eqs:3.4}), that describes a Fermi liquid. 
We use Ma's method of choosing scale dimensions for all relevant quantities and then showing 
self-consistently that these choices lead to a stable fixed point.\cite{Ma_1976} As in 
Sec.\ \ref{sec:II} we assign a scale dimension $[k]=1$ to wave numbers, and $[\omega]=1$ to
frequencies (i.e., we choose a dynamical exponent $z=1$). The latter choice reflects
the linear dispersion relation of the soft modes, see the first term
in Eq.\ (\ref{eq:3.1}), which in a Fermi liquid we do not expect to be changed
by renormalization. We further do not expect the power of wave number
(or frequency) in the Gaussian vertex to be renormalized, and therefore
assign a scale dimension $[q({\bm{k}})]=-(d+1)/2$ and $[q({\bm{x}})]=(d-1)/2$
to the soft field as a function of ${\bm{k}}$ and ${\bm{x}}$, respectively
(i.e., we choose the exponent $\eta$ to be zero). The $P$-propagators, normalized by the
volume, are expected to scale as constants, as they do at Gaussian order, see Eq.\ (\ref{eq:3.2b}). 
We hence assign a scale dimensions $[P({\bm k})] = -d/2$. $\Pslash$ scales differently,
Eq.\ (\ref{eq:3.2c}). This behavior, which again is not expected to change under renormalization,
implies $[\Pslash({\bm k})] = -(d-1)/2$. It is important to stress
that all of these expectations will be verified self-consistently once the RG
scheme is complete, and do {\em not} constitute ad-hoc assumptions. 

With these choices, the $q^2$ term in Eq.\ (\ref{eq:3.1}) is dimensionless; in particular,
$[\gamma]=0$. The constant contribution to the $P^2$ term is also dimensionless,
whereas the $\gamma\omega$ contribution is irrelevant by one power of wave number
or frequency compared to the constant one. We now determine the scale dimensions
of the non-Gaussian terms. For the coupling constants of the cubic terms we find
$[c_{3,0}] = [c_{2,1}] = -(d-1)/2$, $[c_{1,2}] = [c_{0,3}] = -(d+3)/2$, and for those
of the quartic ones, $[c_{4,0}] = [c_{3,1}] = [c_{2,2}] = -(d-1)$, $[c_{1,3}] = [c_{0,4}] = -(d+2)$.
All of the non-Gaussian terms thus have negative scale dimensions for all $d>1$.
It is easy to verify that this is also
true for all terms of higher order in the fields, despite the singular vertices in
${\cal A}_{q-P}$ mentioned above. For instance, the $q^2\Pslash^4$ term, whose
vertex scales as $1/k^3$, has a coupling constant with scale dimension $[c_{2,4}] = -2(d-1)$.
At tree level, the fixed-point action is thus 
given by the Gaussian terms alone, and all others are irrelevant with respect
to the Fermi-liquid fixed point in all dimensions $d>1$. It follows by standard arguments
\cite{Wilson_Kogut_1974} that this remains true order by order in
a loop expansion. All coefficients will in general acquire finite
renormalizations, but the structure of the theory will not change.
An important ingredient in this chain of arguments is the Ward identity
proven in Ref.\ \onlinecite{Belitz_Kirkpatrick_2012}, which identifies
$q$ as a soft mode. This assures that the $q$-vertices
will remain soft under renormalization.

We now consider the particle-particle or Cooper channel. The action is again
schematically given by Eqs.\ (\ref{eq:3.1}, \ref{eq:3.3}, \ref{eqs:3.4}), but with one crucial difference:
The frequency structure (which we have suppressed in our schematic
notation) is different and leads, upon inversion of the quadratic form, to
the characteristic Cooper-ladder structure of the Gaussian $q$-propagator. 
Schematically one obtains, instead of Eq.\ (\ref{eq:3.5}),
\be
\langle q\,q\rangle_{\text{p-p}} = \frac{V}{k + \omega} 
   + \frac{V\gamma_{\text{c}}\,\omega}{(k + \omega)^2}\,\frac{1}{1 + \gamma_{\text{c}}\,\log(1/\omega)}\ ,
\label{eq:3.6}
\ee
where $\gamma_{\text{c}}$ is the interaction amplitude in the Cooper channel. 
As before, we do not distinguish between factors of frequency and factors of
temperature. The structure of the interaction part of Eq.\ (\ref{eq:3.6}) implies
that the $q$-fields in the noninteracting part of the particle-particle sector
of the action must be assigned a different scale dimension than those in the
interacting part, and that the latter is logarithmically irrelevant compared to
the former. We thus conclude that the Fermi-liquid fixed-point action is given by
\be
{\cal A}_{\text{FP}} = \frac{1}{V}\sum_{k,\omega} [k + \omega + \gamma\,\omega]\, q^2 + \frac{1}{V}\sum_{k,\omega} P^2 \ ,
\label{eq:3.7}
\ee
where all channels are included, but $\gamma$ represents the interaction amplitudes
in the particle-hole channel only.\cite{Cooper_channel_footnote} 

We thus have shown that there is a choice of scale dimensions that
makes the Gaussian part of the action, Eq.\ (\ref{eq:3.1}), a stable fixed-point
action, i.e., all other parts of the action are irrelevant with respect to the
fixed point. Furthermore, the fixed point describes a Fermi liquid. While the
arguments leading to this conclusion are deceptively simple, it is important
to realize that their validity relies on two crucial and nontrivial inputs: First,
the Ward identity that guarantees the softness of any $q$-vertices,\cite{Belitz_Kirkpatrick_2012} 
and second, the general loop expansion scheme underlying the renormalization
group, Ref.\ \onlinecite{Wilson_Kogut_1974}.

\subsection{Leading corrections to scaling}
\label{subsec:III.C}

Now consider the least irrelevant operators in Eqs.\ (\ref{eqs:3.4}). These are $c_{3,0}$ and 
$c_{2,1}$ with scale dimensions $-(d-1)/2$, and $c_{4,0}$, $c_{3,1}$, and $c_{2,2}$ 
with scale dimensions $-(d-1)$.
All other terms are more irrelevant by power counting. Furthermore, $c_{3,0}$ and $c_{2,1}$,
which multiply odd powers of the fields, enters all
observables quadratically, and therefore the generic least irrelevant operator $u$ with respect to
the Fermi-liquid fixed point has a scale dimension
\be
[u] = -(d-1)\ ,
\label{eq:3.8}
\ee
where $u$ can stand for either of the least irrelevant $c_{n,m}$ or their appropriate
squares. We note that $c_{2,2}$ is promoted to the same status as $c_{2,1}^2$, despite 
the additional $P$-field, by the singular $1/k$ vertex. This is an important difference
between clean and disordered electrons. In the latter case there are no singular vertices,
and as a result the term of $O(q^2 P^2)$ is more irrelevant than the one of 
$O(q^2 P)$.\cite{Belitz_Kirkpatrick_1997} The same is true of the $\gamma^2\omega^2/k$
vertex in the $q^4$-term, the analog of which is more irrelevant in a disordered system.
As a result, in explicit calculations of leading corrections to Fermi-liquid behavior,
there are structurally distinct terms in the clean case that have no analog in the
disordered case. Examples will be given in Ref.\ \onlinecite{us_tbp}.

The operators collectively denoted by $u$ all become marginal in $d=1$. This indicates the 
instability of the Fermi liquid against the formation of a Luttinger-liquid state. 

\subsection{Derivation of scaling behavior}
\label{subsec:III.D}

We now use the above conclusions to determine the scaling behavior of the observables we are
interested in. Let us first consider the DOS. It is given as an expectation value of
${\bar{\psi}}_{n}\psi_{n}$, which is the massive mode $P_{nn}$ introduced in Sec. \ref{subsec:III.A}.
This couples to the soft mode $q$ via the terms in Eqs.\ (\ref{eqs:3.4}). Keeping in mind
that the $P$-propagator scales as a constant, the DOS can, for scaling purposes, be
be expressed as a series of $q$-correlation functions.\cite{NLsM_footnote}
Schematically, 
\be 
N \sim 1 + \frac{1}{V^2}\sum_{k,\omega}\langle q^2\rangle + \frac{1}{V^4}\sum_{\{k,\omega\}}\langle q^4\rangle + \ldots
\label{eq:3.9} 
\ee 
The RG arguments given above guarantee
that the leading contribution to the DOS correction is given by the
term quadratic in $q$. For the scale dimension of the leading scaling
part of $\delta N$ this implies $[\delta N] = 2[q({\bm{k}})] + 2d = 2[q({\bm x})] = d-1$, which
in turn implies Eqs.\ (\ref{eqs:2.10}). 

Similarly, the static spin susceptibility is given as a $\langle PP\rangle$ correlation function,
and the leading correction to the Fermi-liquid result is given by the terms with coupling
constants $c_{2,1}$ and $c_{2,2}$ in Eqs.\ (\ref{eqs:3.4}). Power counting with the scale
dimensions assigned to the fields in Sec.\ \ref{subsec:III.A} shows that its scale dimension
is also equal to $d-1$. Analogously, the leading corrections to the 
specific-heat coefficient $\gamma_V$ (which can be expressed as an energy-energy
correlation function), and the nematic magnetic susceptibility $\chi_{\text{s,p-wave}}$
are determined by the same terms. We thus have
$[\delta\chis] = [\delta\chi_{\text{s,p-wave}}] = [\delta\gamma_V] = d-1$, which
yields Eqs.\ (\ref{eq:2.23}, \ref{eq:2.25}). Finally, by an analogous argument
we find $[\chiN]=d-1-z=d-2$, which yields Eq.\ (\ref{eq:2.14}).
We thus have derived scaling from the field theory via a RG treatment.

\section{Discussion, and Conclusion}
\label{sec:IV}

We now discuss various aspects of our approach, and of our results.

\subsection{Alternative scaling analyses}
\label{subsec:IV.A}

\subsubsection{Dependence of observables on the least irrelevant operator}
\label{subsubsec:IV.A.1}

In Sec.\ \ref{sec:II} we assigned a scale dimension to the leading
fluctuation corrections to various observables, and used those to derive
their scaling behavior. Alternatively, one can consider the observables
themselves, and use the properties of the Fermi-liquid fixed point,
specifically, the scale dimension of the least irrelevant operator with
respect to it. This line of reasoning has been used in the past for disordered
electrons (see Ref.\ \onlinecite{Belitz_Kirkpatrick_1997} and Appendix \ref{app:B} below) 
and we include it here to show that
it is equivalent to the one given in Sec.\ \ref{subsec:II.E}. We illustrate the
argument by considering the DOS at zero temperature.

At the Fermi-liquid fixed point the DOS is finite, so
we assign it a scale dimension of zero and write, at zero temperature,
\be
N(\omega,u) = N(\omega\,b,u\,b^{-(d-1)})\ .
\label{eq:4.1}
\ee
Here $u$ is the least irrelevant operator in the action, see Eq.\ (\ref{eq:3.8}) and the
accompanying discussion. We now chose $b = 1/\omega$ and obtain
\be  
N(\omega,u) = N(1,u\,\omega^{(d-1)})
\label{eq:4.2}
\ee
Since $N(1,y)$ is evaluated at finite frequency we can Taylor
expand in powers of $y$ with impunity and obtain
\be
N(\omega\to 0) \propto \text{const.} + \vert\omega\vert^{(d-1)}  +\cdots\ ,
\label{eq:4.3}
\ee
which recovers the result from Sec.\ \ref{subsubsec:II.E.1}. This line of reasoning
is closely related to the one used near critical fixed points to obtain corrections
to scaling.\cite{Wilson_Kogut_1974, Cardy_1996}

The same argument can obviously be applied to any other observable. We note,
however, that it hinges on the observable under consideration coupling to one of the
manifestations of $u$. If the leading coupling of some quantity were to, say, $c_{6,0}$ 
that is the coupling constant of the $q^6$ term in Eq.\ (\ref{eq:3.4b}), then the leading 
correction to that quantity would scale as $\omega^{2(d-1)}$, etc.

\subsubsection{Corrections to scaling from the fermionic free energy}
\label{subsubsec:IV.A.2}

An argument related to the one presented in the preceding subsection 
can be used to obtain all of our results in terms of a single
scaling function for the free energy, provided the leading irrelevant variables
are taken into account. Consider Eq.\ (\ref{eq:2.4a}) again, but take into
account the least irrelevant variable $u$, which reflects bosonic fluctuations. 
Keeping only the dependences on $T$ and $H$, and in addition on $u$, we have
\be
f^{(f)} (T,H,u) = b^{-2}\,f^{(f)}(Tb, Hb, u b^{-(d-1)})
\label{eq:4.4}
\ee
Now if we use the fact that $f^{(f)}(1,0,z)$
and $f^{(f)}(0,1,z)$ are analytic functions of $z$. Then we obtain,
for example,
\bse
\label{eqs:4.5}
\be
\gamma_V (T\to 0, H=0) = \text{const.} + c_{\gamma}\,T^{(d-1)}
\label{eq:4.5a}
\ee
and
\be
\chis(T=0, H\to 0) = \text{const.} + c_{\chis}\,H^{(d-1)}\ .
\label{eq:4.5b}
\ee
\ese
Here the $c_{\gamma}$ and $c_{\chis}$ are proportional to $u$ and given
in terms of derivatives of the scaling function.

All other results from Sec.\ \ref{sec:II} can obviously be obtained by an analogous reasoning.
It should be mentioned, however, that this works so easily since the fermionic and bosonic
excitations have the same dynamical scale dimension $z$. If this is not the case, for instance
in the case of disordered electrons, the concept still works but the argument becomes more
complicated.

\subsection{Remarks concerning scaling theories}
\label{subsec:IV.B}

The main point of this paper has been to discuss scaling behavior near the Fermi-liquid
fixed point on various levels of sophistication. We add several comments that complement
the remarks already made in Ref.\ \onlinecite{scaling_footnote}. 

First, scaling works whenever there are soft modes that lead to scale invariance
since processes at long wavelengths and low frequencies dominate the relevant
physics. Crucial questions are, the number and nature of the soft modes, and
the observables and external fields they couple to. The concept of scale invariance
is best known in the context of critical phenomena, where the relevant soft modes
are the critical modes. However, ``generic scale invariance'', which is caused by
soft modes that are due to either Goldstone's theorem or conservation laws, and
which holds in entire phases, is at least equally important, 
see Ref.\ \onlinecite{Belitz_Kirkpatrick_Vojta_2005}. The Fermi-liquid phase provides
a good example of generic scale invariance, with the soft modes in question the
particle-hole excitations, which are Goldstone modes, and the zero-sound and
paramagnon collective modes, which are due to conservation laws.

Second, phenomenological scaling relies on input that informs the scaling
assumptions. This input may be taken from experiment (as was the case in early
studies of critical phenomena), or from theory, which even if incomplete may
provide important clues with respect to the above crucial questions, or from both.
Even if a complete theory is available, simple scaling is still very useful, as it
provides a very simple way to get quick qualitative answers, and to check and
elucidate the physics behind explicit calculations.

Third, in a renormalization-group context, scaling near stable fixed points
that describe entire phases is just as valid and useful as near critical fixed points.
The only difference is the nature of the soft modes (if any; an example of an 
ordered phase without soft
modes is an Ising ferromagnet) that lead to the scale invariance. Moreover, since
stable fixed points tend to be characterized by Gaussian fixed-point Hamiltonians,
{\em exact} results can be obtained for physical dimensions, which is usually not
possible for critical fixed points. Appendix \ref{app:A} provides a very simple
pedagogical example.

Fourth, since all of the observables we have
discussed (except for $\chiN$) have the same naive dimension, they all generically
scale the same way. The only exception is the case of a long-ranged Coulomb
interaction, which leads to a second time scale, the plasmon scale, that acts as
a dangerous irrelevant variable with respect to the DOS and its susceptibility. 
This behavior is less generic than, for instance, the case of critical behavior above
an upper critical dimension, where dangerous irrelevant variables affect all
observables, and naive scaling breaks down. See the next subsection for a discussion
of why the DOS is affected by a long-ranged interaction, whereas other observables
are not.

\subsection{Gauge invariance, and susceptibility of observables to long-range interactions}
\label{subsec:IV.C}

As we have seen in Sec.\ \ref{sec:II}, the nonanalyticities of the DOS and its susceptibility are
sensitive to a long-range Coulomb interaction, whereas those of the specific heat and the
spin susceptibility are not. This can be understood as follows. For fermions interacting
via a Coulomb interaction, the action is invariant under $U(1)$ local gauge transformations,
and in particular under a pure imginary-time transformation $\psi({\bm x},\tau) \to 
\psi({\bm x},\tau)\,\exp{[i\Lambda(\tau)]}$, ${\bar\psi}({\bm x},\tau) \to 
{\bar\psi}({\bm x},\tau)\,\exp{[-i\Lambda(\tau)]}$ with $\psi({\bm x},\tau)$ the fermionic field
as a function of position ${\bm x}$ and imaginary time $\tau$, and ${\bar\psi}$ the adjoint
field. The scalar electromagnetic
potential, which is massless and gives rise to the long-range Coulomb interaction, serves
as the gauge field. The susceptibilities that determine the specific-heat coefficient and the
spin susceptibility are all Fourier transforms of expressions that involve only bilinear
products ${\bar\psi}(\tau)\psi(\tau)$, and hence are gauge invariant. This is not true,
however, for the DOS. If we Fourier transform from the imaginary time variable $\tau$ to
a Matsubara frequency $\omega_n = 2\pi T(n+1/2)$ ($n=0,1,2,\ldots$), and write
$\psi_n \equiv \psi(\omega_n)$, then the DOS is 
related to a product $Q_{nn} = {\bar\psi}_n\psi_n$ that is local in Matsubara
frequency space rather than in imaginary-time space.\cite{Belitz_Kirkpatrick_2012}
For instance, a linear gauge transformation $\Lambda(\tau) = \alpha\tau$ results
in a frequency shift $Q_{nn} \to Q_{n+\alpha,n+\alpha}$. Since the screening of the
Coulomb interaction, which results from integrating out the gauge field, is frequency
dependent, this makes it plausible that the DOS can be sensitive to the difference
between short-ranged and long-ranged interactions. This is also consistent with
the fact that the critical behavior at the metal-insulator transition in disordered
interaction fermion systems depends on the nature of the interaction for the DOS,
but not for gauge invariant quantities.\cite{Finkelstein_1984a, Belitz_Kirkpatrick_1994}

\subsection{The DOS anomaly, and pseudogaps in two-dimensional electron systems}
\label{subsec:IV.D}

The linear frequency or energy dependence of the DOS for $d=2$, Eqs.\ (\ref{eq:2.10b}, \ref{eq:2.12}),
is of interest in the context of the ``pseudogap'' feature of the DOS that is observed in many
strongly correlated $2-d$ electron systems. This feature, which was first discussed by
Mott\cite{Mott_1968} in the context of generic strongly correlated sytems, later became
strongly associated with high-$T_{\text{c}}$ superconductivity, and the superconducting
gap in these materials is widely believed to develop out of the pseudogap.\cite{Timusk_Statt_1999}
However, a recent experiment casts doubt on this notion in at least some
superconductors.\cite{Jacobs_et_al_2012} It is thus possible that at least some of the
observed pseudogaps reflect a generic feature of a Fermi liquid, viz., the leading
correction to scaling for the DOS, rather than being a harbinger of superconductivity.
They still reflect a ``strange-metal''-aspect of strongly correlated electrons, however:
Since the DOS is the order parameter for the Fermi liquid, strong correlations can
induce a quantum phase transition to a non-Fermi-liquid phase where the DOS at
the Fermi surface vanishes,\cite{Kirkpatrick_Belitz_2012} and the DOS anomaly in the Fermi-liquid phase is a
precursor of this transition.

\subsection{The density-of-states fluctuations}
\label{subsec:IV.E}

To illustrate how strong the effects of the Goldstone modes are on
the DOS susceptibility, Sec.\ \ref{subsubsec:II.E.2}, let us use simple statistical
arguments to determine the behavior of $\chi_{\text{OP}}$ one would
expect in the absence of anomalous fluctuations. 
Consider $\varphi_{n}({\bm{x}})=\psi_{n}({\bm{x}})/\sqrt{T}$, and 
$p_n({\bm x}) = {\bar{\varphi}}_{n}({\bm{x}})\,\varphi_{n}({\bm x})$, and
define the ``volume'' $V_{T}\equiv1/T$ in the imaginary-time direction
of the space-time of quantum statistical mechanics. Then 
$\langle p_{n}({\bm{x}})\rangle \propto V_{T}$
is a ``time-extensive'' quantity that is proportional to $V_{T}\equiv1/T$.
Now consider the flucutation $\langle(\delta p_n({\bm x}))^2\rangle$, the
connected part of which is proportional to $1/T = V_T$, see Appendix \ref{app:D}.
For the relative fluctuation this implies $\langle(\delta p_{n}({\bm{x}}))^{2}\rangle/
\langle p_n ({\bm{x}})\rangle^{2}\propto V_{T}/V_{T}^{2}=1/V_{T}=T$.
This just says that the relative fluctuation is proportional to $1/V_{T}$,
as one would expect from ordinary statistics. This yields an estimate
for the fluctuations of $\rho({\bm x},i\omega_n) = {\bar\psi}_n({\bm x})\,\psi_n({\bm x})$
which determine $\chiN$, see Eq.\ (\ref{eq:D.1}): 
\bea
\langle(\delta{\rho}({\bm x},i\omega_n))^2\rangle &\propto& \frac{\langle(\delta p_n)^2\rangle}{V_T^2}
\propto \frac{\langle(\delta p_n)^2\rangle/V_T^2}{(\langle p_n\rangle/V_T)^2}
\nonumber\\
&\propto& 1/V_T = T\ . 
\label{eq:4.6} 
\eea 
These arguments assume that
there are no strong fluctuations in the system that invalidate the
simple statistics. For the connected part of $\chiN$, which we
denoted by $\chiN^{(1)}$ in Sec. \ \ref{sec:II}, we
thus have 
\be \chiN^{(1)}({\bm k},i\omega_n;T) = T\,\theta({\bm k},i\omega_n;T)\ . 
\label{eq:4.7} 
\ee
In the absence of anomalous
fluctuations, $\theta$ will scale as the zeroth power of the wave
number, the frequency, or the temperature; i.e., $\theta\sim1$.

We conclude that if the DOS were normally distributed, we would have
$\chiN^{(1)} = T\times O(1)$. From Eq.\ (\ref{eq:2.16b})
we know that this is not correct even in the case of
a short-ranged interaction. Instead, the quantity $\theta$ in Eq.\ (\ref{eq:4.7})
scales as $\theta\sim1/k\sim1/T$ in $d=2$, and $\chiN^{(1)}\sim1$. This
divergence of the relative DOS fluctuations reflects the strong fluctuations
in the system that are a consequence of the existence of the Goldstone
modes. In $d=3$ the fluctuations are weaker, and the dependence of
$\theta$ on $k$ or $T$ is only logarithmic, see Eq.\ (\ref{eq:2.18b}).

As we have seen in Sec.\ \ref{sec:II}, a long-ranged Coulomb interaction further
amplifies these effects. Equation (\ref{eq:2.21}) shows that in $d=2$, $\theta \sim 1/k^{3/2}$.
Even more remarkable is the fact that in both $d=2$ and $d=3$, $\chiN$ diverges
in the limit of a vanishing wave number $k\to 0$, see Eqs.\ (\ref{eq:2.21}) and (\ref{eq:2.22}).
Putting $k=0$ and considering a finite system with linear dimension $L$ we have,
at any nonzero temperature,
\be
\chi_N \propto \begin{cases} \ln L & \text{for $d=2$}\\
                                                L      & \text{for $d=3$}\ .
                         \end{cases}
\label{eq:4.8}
\ee
The reason for this unusual behavior is the breakdown of screening 
of the Coulomb interaction at nonzero frequencies. $\chiN$ is
susceptible to both the effects of the Goldstone modes and the
breakdown of screening. The effects of the former are stronger
in $d=2$ than in $d=3$, whereas for the latter the opposite is true.
This raises the following interesting point. Consider the quantity
$p_n({\bm x})$ as defined above, which is subject to thermal and
quantum fluctuations that are described by a probability density
function $P$. The expected value, $\langle p\rangle = \int D[p]\,p\,P[p]$,
exists and determines the density of states. However, the second moment,
which determines $\chiN$, does not exist in either $d=2$ or $d=3$,
and it is easy to see that none of the higher moments exist either.
The density of states therefore must have a broad distribution
that cannot be represented by a Gaussian. This phenomenon
requires a separate investigation.

Comparing the results in Sec.\ \ref{subsubsec:II.E.2} with the
Fermi-liquid result for $\chiN$ we see that the bosonic contributions
give the {\em leading} behavior of $\chiN$ in $d\leq 2$. This is
 in contrast to all other quantities, where the latter give a
{\em correction} to the Fermi-liquid result for all $d>1$. This is because that $\chiN$, as the OP susceptibility,
couples particularly strongly to the Goldstone modes. This is precisely
analogous to the OP susceptibility in a classical Heisenberg ferromagnet,
whose leading behavior in $d\leq4$ is also determined by the coupling to the
Goldstone modes, see Appendix \ref{app:A}.

We also add some comments about the experimental relevance of the
quantity $\chiN$. In any system a local measurement of the DOS depends
on the position and is referred to as the local density of states 
(LDOS). The LDOS gives the dominant contribution to the tunneling current in 
a scanning tunneling microscope.\cite{Tersoff_Hamann_1985} Its
average is the DOS as calculated in the present paper and also measured in
a tunnel junction. Our OP susceptibility, Eq.\ (\ref{eq:D.1}), describes the 
averaged fluctuations of the LDOS. A suitable two-tip tunneling experiment 
should be able to give information about this quantity.

\subsection{Conclusion, and Outlook}
\label{subsec:IV.F}

In summary, we have presented a scaling analysis of nonanalyticities in Fermi
liquids. The most important conclusion is that the exponents of various
nonanalyticities that were first derived in perturbation theory are exact. In
addition, the scaling theory allows for a unified treatment of clean and
disordered electronic systems, as well as various analogous phenomena
in classical many-body systems. This demonstrates the generality of the
method, which can also be applied to more exotic conductors, such as
Dirac and Weyl metals.

\acknowledgments 
We gratefully acknowledge helpful correspondence and discussions
with Andrey Chubukov, S. Gregory, H. Manoharan, and Dmitrii Maslov.
This work was supported by the National Science Foundation
under Grant Nos. DMR-09-29966 and DMR-09-01907. Part
of this work was performed at the Aspen Center for Physics and supported
by the NSF under Grant No. PHYS-1066293. We thank the Center for its
hospitality.

\appendix

\section{A simple example: Scaling analysis of $\phi^{4}$-theory}
\label{app:A}

It is illustrative to recall the scaling analysis for classical ferromagnets that
is analogous to our treatment of the clean fermion action in Sec.\ \ref{sec:III}.
Consider an $O(2)$ $\phi^4$-theory with a $2$-component field 
${\bm\phi}({\bm x}) = \left(\phi_1({\bm x}),\phi_2({\bm x})\right)$ and an action
\be
S = \int d{\bm x}\,\left[\frac{r}{2}\,{\bm\phi}^2({\bm x}) + \frac{c}{2}\,\left(\nabla{\bm\phi}({\bm x})\right)^2 + \frac{u}{4}\left({\bm\phi}^2({\bm x})\right)^2\right]\ .
\label{eq:A.1}
\ee
The properties of the ordered phase are usually described by parameterizing
${\bm\phi}({\bm x}) = \rho({\bm x}){\hat{\bm\phi}}({\bm x})$, with ${\hat{\bm\phi}}$ 
a unit vector.\cite{Zinn-Justin_1996} Here we deliberately choose a different
parameterization in order to to illustrate our treatment of the fermion problem
in the context of a much simpler model.

A saddle-point solution corresponding to the ordered phase is
${\bm\phi}_{\text{sp}}({\bm x}) = (\phi_0,0)$ with $\phi_0 = \sqrt{-r/u}$.
Now write $\phi_1({\bm x}) = \phi_0\left(1 + p({\bm x})\right)$ and
$\phi_2({\bm x}) = \phi_0 \pi({\bm x})$, and expand in the fluctuations
$p$ and $\pi$. Then we obtain a Gaussian action
\be
S^{(2)} = \frac{1}{V}\sum_k k^2\,\pi^2 + \frac{1}{V}\sum_k [1 + O(k^2)] \, p^2\ .
\label{eq:A.2}
\ee
Here we have performed a Fourier transform from $\pi({\bm x})$ and
$p({\bm x})$ to $\pi({\bm k}) \equiv \pi$ and $p({\bm k}) \equiv p$,
we have rescaled the fields to make the Gaussian coupling constant
equal to unity, and we use the same schematic notation as in Sec.\ \ref{sec:III}.
The non-Gaussian part of the action takes the form
\be
\Delta S = \Delta S^{(3)} + \Delta S^{(4)}
\label{eq:A.3}
\ee
where
\bse
\label{eqs:A.4}
\bea
\Delta S^{(3)} &=& \frac{c_{2,1}}{V^2} \sum_{\{k\}} \pi^2\,p + \frac{c_{0,3}}{V^2}\sum_{\{k\}} p^3\ ,
\label{eq:A.4a}\\
\Delta S^{(4)} &=& \frac{c_{4,0}}{V^3} \sum_{\{k\}} \pi^4 + \frac{c_{2,2}}{V^3} \sum_{\{k\}} \pi^2\,p^2 + \frac{c_{0,4}}{V^3} \sum_{\{k\}} p^4\ .
\nonumber\\
\label{eq:A.4b}
\eea
\ese
The bare values of the coupling constants $c_{n,m}$ can be expressed in terms of the coupling
constants in the original action, Eq.\ (\ref{eq:A.1}). 

Under renormalization, terms of higher order
in the fields are generated, and the coupling constants acquire a wave-number dependence.
By symmetry the latter takes the form of a dependence on $k^2$. Furthermore, the
$O(2)$-symmetry of the action leads to a Ward identity that guarantees that the
transverse fluctuation $\pi$ is a soft mode.\cite{Zinn-Justin_1996} This is correctly
reflected in the Gaussian action, Eq.\ (\ref{eq:A.2}). However, $\Delta S$ contains
terms where $\pi$ appears without any gradients. The Ward identity ensures that
the zeroth-order contributions to these terms in a gradient expansion cancel, and
for power-counting purposes $c_{0,4}$, for instance, must be written as 
\be
c_{4,0} = {\tilde c}_{0,4} k^2\ ,
\label{eq:A.5}
\ee
and analogously for $c_{2,1}$ and $c_{2,2}$. Indeed, explicitly integrating out $p$
shows that at tree level the term proportional to $c_{2,1}^2$ cancels the term
proportional to $c_{4,0}$. 

We now assign scale dimensions in an attempt to find a stable fixed point that
describes the ferromagnetic phase. We know that $\pi$ is soft, and that the 
Goldstone modes are proportional to $k^2$, which means the first term in
Eq.\ (\ref{eq:A.2}) must be part of the fixed-point action. We also know that
the $p$-correlations are short-ranged, which implies that the second term
in Eq.\ (\ref{eq:A.2}) is part of the fixed-point action as well. With $[k] = 1$
the scale dimension of the wave number as in Sec.\ \ref{sec:III}, this motivates
$[\pi] = -(d+2)/2$ and $[p] = -d/2$. Power counting then shows that all
terms in Eqs.\ (\ref{eqs:A.4}) are irrelevant. The fixed-point action is thus
given by Eq.\ (\ref{eq:A.2}), and the least irrelevant operator is
${\tilde c}_{4,0}$ with $[{\tilde c}_{4,0}] = -(d-2)$. This implies that $d=2$ is a lower
critical dimension for the problem, consistent with the Mermin-Wagner
theorem.

We next add an external magnetic field $h$ in the $\phi_1$-direction to the problem. 
By shifting $p$ one sees that the leading coupling of the field to the soft mode
takes the form $h\pi^2$, which gives $h$ a scale dimension $[h]=2$. 
Now consider the normalized magnetization, $m = \langle\sqrt{1-\pi^2({\bm x})}\rangle$, which
is the order parameter of the system. The leading fluctuation correction to $m$
is thus given by the correlation function $\delta m = \langle \pi({\bm x})\,\pi({\bm x})\rangle$,
whose scale dimension is $[\delta m] = d-2$. The relevant homogeneity law is thus
\be
\delta m(h) = b^{-(d-2)} \delta m(h b^2)\ ,
\label{eq:A.6}
\ee
which yields
\be
m \propto {\text{const.}} + h^{(d-2)/2}\ .
\label{eq:A.7}
\ee
This nonanalytic field-dependence of the magnetization\cite{Ma_1976} is a result of the
Goldstone modes, i.e., the ferromagnons, that are represented by the soft
$\pi$-fluctuations. An equivalent manifestation of the Goldstone modes is
the behavior of the longitudinal susceptibility, $\chi_{\text{L}} = \partial m/\partial h$,
which diverges for $h\to 0$ for all $d<4$:\cite{Brezin_Wallace_1973}
\bse
\label{eqs:A.8}
\be
\chi_{\text{L}}(h) \propto h^{-(d-4)/2}\ .
\label{eq:A.8a}
\ee
Alternatively, the zero-field inhomogenous susceptibility diverges for small wavenumbers as
\be
\chi_{\text{L}}(k) \propto k^{-(4-d)}\ .
\label{eq;A.8b}
\ee
\ese
Note the close analogy between these results for the magnetic order parameter and
its susceptibility, and those for the DOS, which is the order parameter
for the Fermi-liquid state, and its susceptibility in Sec. \ \ref{subsec:II.E}.

We finally mention that one can integrate out $p$ in a saddle-point approximation
that keeps $\pi$ fixed. For $p$ as defined above Eq.\ (\ref{eq:A.2}) (before the scaling
that normalized the coefficients in the Gaussian action) this leads to 
\bea
p({\bm x}) &=& \sqrt{1-\pi^2({\bm x})} - 1 + \frac{c/2u\phi_0^2}{1-\pi^2({\bm x})}{\bm\nabla}^2 \sqrt{1-\pi^2({\bm x})} 
\nonumber\\
&&\hskip 120 pt + O(\nabla^4)\ .
\label{eq:A.9}
\eea
Substituting this solution of the saddle-point equation back into the action leads
to the familiar nonlinear sigma model
\bea
S_{\text{NL$\sigma$M}} &=& \frac{c}{2}\,\phi_0^2 \int d{\bm x}\ \left[\left({\bm\nabla}\pi({\bm x})\right)^2 + \left({\bm\nabla}\sqrt{1-\pi^2({\bm x})}\right)^2\right]
\nonumber\\
&&\hskip 120pt + O(\nabla^4)\ .
\label{eq:A.10}
\eea
This derivation of the nonlinear sigma model, which provides an alternative to the
usual derivation based on rotational symmetry,\cite{Zinn-Justin_1996} is the
$O(2)$ equivalent of the derivation of an effective action for clean electrons entirely in terms of the
soft $q$-field in Ref.\ \onlinecite{Belitz_Kirkpatrick_2012}. In the electron case, however,
the result is {\em not} a sigma model, and it has not been formulated in a closed form.
This is partly due to the more complicated structure of the vertices in the electronic
model.

\section{Soft modes, scaling, and nonanalyticities in disordered electron systems}
\label{app:B}

In this appendix we recall some features of the disordered electron problem, to the
extent that they are helpful in understanding the corresponding properties of clean
systems discussed in the present paper. 

The soft-mode effective theory for noninteracting disordered electrons does take the form of a
matrix nonlinear sigma model,\cite{Wegner_1979} and its generalization to
interacting systems adds extra terms to the sigma model.\cite{Finkelstein_1983}
After the analog of the massive field $P$ in Sec.\ \ref{sec:III} has been integrated
out, the structure of the model is, in the same schematic notation as in 
Sec.\ \ref{sec:III},\cite{Belitz_Kirkpatrick_1997}
\be
{\cal A}_{\text{NL$\sigma$M}} = \frac{1}{V}\sum_{k,\omega} \left[k^2/G + H\omega + \gamma\omega\right] q^2 + O(k^2\,q^4, \omega\,q^3)\ .
\label{eq:B.1}
\ee
Here $q$ represents the soft components of bilinear fermion fields as in Sec.\ \ref{sec:III}, and
the nonlinear-sigma-model part of the theory has been expanded in powers of $q$, keeping
only the quadratic term. There are several important differences between this model and
the clean model of Ref.\ \onlinecite{Belitz_Kirkpatrick_2012}, despite their apparent similarity. One is that in the disordered case,
various observables appear as coupling constants of the field theory. $G$ in Eq.\ (\ref{eq:B.1})
is proportional to the electrical resistivity, $H$ is proportional to the specific-heat
coefficient, and $H$ plus the spin-singlet and spin-triplet interaction constants 
summarily denoted by $\gamma$ in Eq.\ (\ref{eq:B.1}) determine the density and spin susceptibilities,
respectively.\cite{Belitz_Kirkpatrick_1994} In contrast, the corresponding observables 
in the clean case need to be calculated as correlation functions of the basic matrix field.
Partly as a result of that, the fixed point describing the disordered Fermi liquid is easier
to obtain than in the clean case. Let us assign a scale dimension
$[q({\bm k})] \equiv [q] = -(d+2)/2$ to the matrix field, and a dynamical exponent
$[\omega] = z = 2$ to the frequency. $G$, $H$, and $\gamma$ are then all dimensionless,
and the only term shown explicitly in Eq.\ (\ref{eq:B.1}) represents the fixed-point action. It
describes the diffusive modes in a fermion system with quenched disorder. 
The least irrelevant operators with respect to this fixed point,  which we collectively denote by $u$,  
all have scale dimensions $[u] = -(d-2)$. This suffices to determine the leading
scaling behavior of various observables. For instance, the electrical conductivity $\sigma$
and the specific-heat coefficient $\gamma_V$ are both dimensionless according to the
above arguments. $\sigma$ thus obeys a homogeneity law
\bse
\label{eqs:B.2}
\be
\sigma(\omega,u) = \sigma(\omega b^2, ub^{-(d-2)})
\label{eq:B.2a}
\ee
which results in a low-frequency nonanalyticity or long-time tail
\be
\sigma(\omega\to 0) \propto \text{const.} + \omega^{(d-2)/2}\ .
\label{eq:B.2b}
\ee
\ese
Similarly, the specific-heat coefficient obeys
\bse
\label{eqs:B.3}
\be
\gamma_V(T,u) =\gamma_V(T b^2, u b^{-(d-2)})
\label{eq:B.3a}
\ee
which results in
\be
\gamma_V(T\to 0) \propto \text{const.} + T^{(d-2)/2}\ .
\label{eq:B.3b}
\ee
\ese
The scaling behavior of the DOS and the spin susceptibility can be obtained
by analogous arguments. All of these results, which are
analogous to the ones for clean systems derived in Secs.\ \ref{sec:III} and \ref{subsec:IV.A}
above, were first derived in perturbation theory.\cite{Altshuler_Aronov_1984} Arguments
analogous to those put forward in Sec.\ \ref{subsec:III.A} later showed that they
represent the {\em exact} (as far as the exponents are concerned) leading nonanalyticities.\cite{Belitz_Kirkpatrick_1997}
Also note that these scaling arguments immediately show that $d=2$ is a lower
critical dimensionality of the problem, as the disordered Fermi-liquid fixed point
becomes unstable for $d\leq 2$.

\section{Long-time tails in classical fluids}
\label{app:C}

Here we sketch how long-time tails in classical fluids can be considered as corrections to
scaling at a Navier-Stokes fixed point. Focussing on the viscosity, we first
derive the long-time tail as deriving from the dependence of the viscosity on the
least irrelevant operator, in analogy to the treatment of clean electrons in Sec.\ \ref{subsec:IV.A},
and of disordered ones in the previous appendix. We then show how, alternatively, it can be
understood as the leading scaling behavior of the fluctuation correction to the viscosity, in
analogy to the development in Sec.\ \ref{sec:II}. Our goal is to 
demonstrate how universally
useful and applicable the notion of corrections to scaling near a stable fixed point is, and
that it applies to classical many-body systems as well as to quantum ones.

\subsection{Corrections to scaling at a Navier-Stokes fixed point}
\label{subapp:C.1}

We focus what arguable is the simplest example of a classical long-time tail, viz., the
one related to the kinematic viscosity $\nu$. For simplicity, we consider incompressible
flow (which in particular eliminates sound waves), and we neglect the pressure-gradient
term. The Langevin equation for the transverse fluid velocity ${\bm u}$ then reads\cite{Landau_Lifshitz_VI_1987}
\be
\partial_t{\bm u} + ({\bm u}\cdot{\bm\nabla}){\bm u} = \nu_0{\bm\nabla}^2{\bm u} + {\tilde{\bm F}}\ .
\label{eq:C.1}
\ee
Here ${\tilde{\bm F}}$ is a Gaussian distributed random force whose second moment is fixed
by the requirement that Eq.\ (\ref{eq:C.1}) correctly render the equilibrium velocity fluctuations:
\bea
\langle{\tilde F}_i({\bm x},t)\,{\tilde F}_j({\bm x}',t')\rangle &\equiv& G_{ij}({\bm x},t\vert{\bm x}',t') 
\nonumber\\
       &=& 2T\nu_0\partial_i\partial_j\,\delta({\bm x}-{\bm x}')\,\delta(t-t')\ .\qquad
\label{eq:C.2}
\eea
$\nu_0$ is the bare kinematic viscosity, which gets renormalized to the physical one $\nu$
by the nonlinear term in Eq.\ (\ref{eq:C.1}). We now show how to use renormalization-group 
and scaling arguments to determine this renormalization. This can be done by using a
Martin-Siggia-Rose formalism\cite{Martin_Siggia_Rose_1973, Bausch_Janssen_Wagner_1976, DeDominicis_Peliti_1978} 
to cast the problem in a field-theoretic language.

We start with the generating functional for all correlation functions of the fluid velocity:
\begin{widetext}
\be
Z[{\tilde{\bm F}}] = \int D[{\bm u}]\,\delta\left[\partial_t{\bm u} + ({\bm u}\cdot{\bm\nabla}){\bm u} - \nu_0{\bm\nabla}^2{\bm u} - {\tilde{\bm F}}\right]\ 
e^{-\frac{1}{2}\int d{\bm x}\,d{\bm x}'\,dt\,dt'{\tilde F}_i({\bm x},t)\, G_{ij}^{-1}({\bm x},t\vert{\bm x}',t')\, {\tilde F}_j({\bm x}',t')}
\label{eq:C.3}
\ee
Here $D[{\bm u}]$ is a functional integration measure.
Enforcing the functional delta-constraint by means of an auxiliary field ${\bar{\bm u}}$, and integrating
out the Langevin noise, we obtain
\bse
\label{eqs:C.4}
\be
Z \equiv \int D[{\tilde{\bm F}}]\,Z[{\tilde{\bm F}}] = \int D[{\bm u},{\bar{\bm u}}]\ e^{-S[{\bm u},{\bar{\bm u}}]}
\label{eq:C.4a}
\ee
with an action
\be
S[{\bm u},{\bar{\bm u}}] = i\int d{\bm x}\,dt\,{\bar{\bm u}}\cdot\left[\partial_t{\bm u} + c\,({\bm u}\cdot{\bm\nabla}){\bm u} - \nu_0{\bm\nabla}^2{\bm u}\right]
    + \frac{1}{2}\int d{\bm x}\,{\bm x}'\,dt\,dt'\,{\bar u}_i({\bm x},t)\, G_{ij}({\bm x},t\vert{\bm x}',t')\, {\bar u}_j({\bm x}',t')
\label{eq:C.4b}
\ee
\ese
\end{widetext}
Here we have introduced a nominal coupling constant $c$ for the nonlinear term, whose bare value is $c_0=1$.

Now we assign scale dimensions $[L]=-1$ and $[t]=-z$ to length and time, respectively. Choosing
\bse
\label{eqs:C.5}
\be
[{\bm u}] = [{\bar{\bm u}}] = d/2
\label{eq:C.5a}
\ee
and 
\be
z = 2
\label{eq:C.5b}
\ee
\ese
leads to a stable Navier-Stokes fixed point with respect to which the nonlinear term is irrelevant. Indeed, the
scale dimension of the coupling constant $c$ is negative for $d>2$,
\be
[c] = -(d-2)/2\ .
\label{eq:C.6}
\ee
while $\nu_0$ is dimensionless and the viscosity term is thus part of the fixed-point action.
Since the nonlinear term is cubic in the fields, it always appears squared in explicit calculations. The wave-number
and frequency dependent kinematic viscosity thus obeys a homogeneity law
\be
\nu(k,\omega) = f_{\nu} (kb,\omega b^2,c^2 b^{-(d-2)})
\label{eq:C.7}
\ee
with $f_{\nu}$ a scaling function. From Eq.\ (\ref{eq:C.7}) we obtain in particular, by expanding in powers of the small
third argument,
\be
\nu(k=0,\omega) \propto 1 + \text{const.}\times \omega^{(d-2)/2}\ ,
\label{eq:C.8}
\ee
where the constant is proportional to $c^2$. 
This is the well-known classical long-time tail.\cite{Pomeau_Resibois_1975} In $d=2$ 
$c$ becomes marginal, which reflects the fact that the local description of hydrodynamics
breaks down in $d\leq 2$.\cite{Pomeau_Resibois_1975, Forster_Nelson_Stephen_1977}

\subsection{Scaling of the fluctuation correction}
\label{subapp:C.2}

To complete the analogy with our various discussions of the quantum problem, we now consider the
long-time tail of the classical viscosity from an alternative point of view. The kinematic viscosity $\nu$
is defined as the shear viscosity $\eta$ divided by the mass density $\rho$, $\nu = \eta/\rho$, and thus
has an naive dimension of a length squared divided by a time. With the choice of scale dimensions
specified in Sec. \ \ref{subapp:C.1}, this makes $\nu$ dimensionless, consistent with the fixed-point
action identified above. Now consider the leading correction $\delta\nu$ to the kinematic viscosity.
To determine the scale dimension of $\delta\nu$, we recall that the viscosity physically results from
a frictional force ${\bm F}_f$. With $G$ a friction coefficient, we write
\be
{\bm F}_f = G{\bm u}\ .
\label{eq:C.9}
\ee
Considering planar Couette flow within a hypercube of linear dimension $L$ we have, for the $x$-component $F_f$
of ${\bm F}_f$, and with an accuracy that suffices for dimensional arguments,
\be
F_f = G L \partial u_x/\partial y\ .
\label{eq:C.10}
\ee
On the other hand, the $x$-$y$ component of the stress tensor $T$ is given by
\be
T_{xy} = \eta \partial u_x/\partial y\ .
\label{eq:C.11}
\ee
To obtain the frictional force we need to multiply by the cross-sectional area $L^{d-1}$. Equating
the result with Eq.\ (\ref{eq:C.10}) we obtain
\be
G L \partial u_x/\partial y = L^{d-1} \eta\,\partial u_x/\partial y\ ,
\label{eq:C.12}
\ee
and finally
\be
G = L^{d-2}\eta\ .
\label{eq:C.13}
\ee
$G/\rho$ thus scales as $L^{d-2}$ times a dimensionless quantity, which yields $[\delta\nu] = -(d-2)$
for the scale dimension of $\delta\nu$. The appropriate homogeneity law, according to Eq.\ (\ref{eq:2.1}),
is thus 
\be
\delta\nu(k,\omega) = b^{(d-2)}\delta\nu(kb,\omega b^2)\ .
\label{eq:C.14}
\ee
Equation (\ref{eq:C.8}) now follows as the leading scaling behavior of $\delta\nu$. 

The above arguments, which are physically equivalent to the ones given in Appendix \ref{subapp:C.1}, are
analogous to the scaling theory of electron localization by Abrahams et al.\cite{Abrahams_et_al_1979}
Building on arguments by Thouless, these authors realized that the natural scaling variable is the
conductance $G$, which is related to the conductivity $\sigma$ by $G = L^{d-2}\sigma$. The relation
between the friction coefficient $G$ and the viscosity $\eta$ in Eq.\ (\ref{eq:C.13}) is the precise analog
for the classic fluid case.

\section{The structure of the density-of-states susceptibility}
\label{app:D}

The density-of-states susceptibility $\chiN$ is given as a four-fermion correlation
function\cite{Kirkpatrick_Belitz_2013}
\be
\chiN({\bm x}-{\bm y};i\omega_n,i\omega_m) = \langle\delta\rho({\bm x},i\omega_n)\,
     \delta\rho({\bm y},i\omega_m)\rangle
\label{eq:D.1}
\ee
Here $\rho({\bm x},i\omega_n) = {\bar\psi}_n({\bm x})\,\psi_n({\bm x})$, and 
$\delta\rho = \rho - \langle\rho\rangle$. Defining a Fourier transform from
a Matsubara frequency $\omega_n$ to an imaginary time $\tau$ as in 
Ref.\ \onlinecite{Kirkpatrick_Belitz_2013} this takes the form
\begin{widetext}
\be
\chiN({\bm x}-{\bm y};i\omega_n,i\omega_m) = T^2 \int_0^{1/T} d\tau_1\,d\tau_2\,d\tau_3\,d\tau_4
   \ e^{i\omega_n(\tau_1-\tau_2) + i\omega_m(\tau_3-\tau_4)} \langle\delta\left({\bar\psi}({\bm x},\tau_1)\,\psi({\bm x},\tau_2)\right)\,\delta\left({\bar\psi}({\bm y},\tau_3)\,\psi({\bm y},\tau_4)\right)\rangle
\label{eq:D.2}
\ee
There are two distinct contributions to this correlation function: First, a disconnected one
in which the four-fermion correlation factorizes into a product of two two-fermion correlations:
\bse
\label{eqs:D.3}
\be
\chiN^{\text{dc}}({\bm x}-{\bm y};i\omega_n,i\omega_m) = -T^2 \int_0^{1/T} d\tau_1\,d\tau_2\,d\tau_3\,d\tau_4
   \ e^{i\omega_n(\tau_1-\tau_2) + i\omega_m(\tau_3-\tau_4)} \langle{\bar\psi}({\bm x},\tau_1)\psi({\bm y},\tau_4)\rangle \langle{\bar\psi}({\bm y},\tau_3)\psi({\bm x},\tau_2)\rangle \ ,
\label{eq:D.3a}
\ee
and second, a connected one that contains the contributions to
$\langle{\bar\psi}\psi{\bar\psi}\psi\rangle$ that do not factorize:
\be
\chiN^{\text{c}}({\bm x}-{\bm y};i\omega_n,i\omega_m) = T^2 \int_0^{1/T} d\tau_1\,d\tau_2\,d\tau_3\,d\tau_4
   \ e^{i\omega_n(\tau_1-\tau_2) + i\omega_m(\tau_3-\tau_4)} \langle{\bar\psi}({\bm x},\tau_1)\psi({\bm x},\tau_2){\bar\psi}({\bm y},\tau_3)\psi({\bm y},\tau_4)\rangle^{\text{c}}\ .
\label{eq:D.3b}
\ee
\ese
Using time translational invariance, we finally obtain
\bse
\label{eqs:D.4}
\be
\chiN^{\text{dc}}({\bm x}-{\bm y};i\omega_n,i\omega_m) = -\delta_{nm} \int_0^{1/T} d\tau\,d\tau'\,
     {\cal G}({\bm x}-{\bm y},\tau)\,{\cal G}({\bm y}-{\bm x},\tau')\ ,
\label{eq:D.4a}
\ee
where ${\cal G}({\bm x},\tau) = \langle{\bar\psi}({\bm x},\tau)\,\psi(0,0)\rangle$, and
\be
\chiN^{\text{c}}({\bm x}-{\bm y};i\omega_n,i\omega_m) = T \int_0^{1/T} d\tau_1\,d\tau_2\,d\tau_3\,
   \ e^{i\omega_n(\tau_1-\tau_2) + i\omega_m(\tau_3)} \langle{\bar\psi}({\bm x},\tau_1)\psi({\bm x},\tau_2){\bar\psi}({\bm y},\tau_3)\psi({\bm y},0)\rangle^{\text{c}}\ .
\label{eq:D.4b}
\ee
\ese
\end{widetext}
This shows that $\chiN^{\text{dc}}$ approaches a constant as $T\to 0$, whereas $\chiN^{\text{c}}$ is proportional to $T$. These 
are the properties we used in Sec.\ \ref{sec:II}, where we denoted $\chiN^{\text{dc}}$ and $\chiN^{\text{c}}$ by $\chiN^{(0)}$
and $\chiN^{(1)}$ respectively.


\begin{thebibliography}{66}
\bibitem[{not()}]{notation_footnote}
\bibinfo{note}{We write $a\sim b$ for `$a$ scales as $b$', and $a\propto b$ for
  `$a$ is proportional to $b$'.}
\bibitem[{sca({\natexlab{b}})}]{scale_dimension_footnote}
\bibinfo{note}{For instance, the density of states $N$ is an inverse energy
  times an inverse volume, so $[N] = - [E] - [V] = d-z$, etc. There is some
  ambiguity here, since one can choose a different value for $[A]$, which is
  compensated by the scale dimension of some variable $A$ depends on. However,
  using the naive scale dimension is arguably the easiest and most transparent
  choice, and we adopt this convention.}
\bibitem[{hig()}]{higher_loop_footnote}
\bibinfo{note}{Whether at two-loop or higher order a $\vert\omega\vert$
  behavior with a nonzero prefactor emerges is an open question.}
\end{thebibliography}

\end{document}